\documentclass[aps,prx,reprint,groupedaddress,superscriptaddress,floatfix]{revtex4-2}

\usepackage[T1]{fontenc}
\usepackage{newtxtext}
\usepackage{newtxmath}
\usepackage{bm}

\usepackage{titlesec}

\usepackage{amsmath}
\usepackage{dsfont}
\usepackage{mathtools}
\usepackage{amsfonts}
\usepackage{bbold}
\usepackage{physics}

\usepackage{color}
\usepackage{xcolor}
\usepackage{soul}
\usepackage[colorlinks,citecolor=blue,linkcolor=blue,urlcolor=blue]{hyperref}

\usepackage{graphicx}
\usepackage[all]{hypcap}

\usepackage{enumitem}

\usepackage{tabularx}
\usepackage{makecell}

\graphicspath{{../img/}}

\usepackage{hyperref}
\usepackage{lineno}
\usepackage{cleveref}
\usepackage{etoolbox}

\makeatletter
\let\addcontentslineOriginal\addcontentsline
\let\addcontentsline\@gobblethree
\makeatother

\begin{document}
\title{Floquet Recurrences in the Double Kicked Top}

\author{Avadhut V. Purohit}
\email{avdhoot.purohit@gmail.com}
\affiliation{Department of Physics, Visvesvaraya National Institute of Technology, Nagpur 440010, India}
\author{Udaysinh T. Bhosale}
\email{udaysinhbhosale@phy.vnit.ac.in}
\affiliation{Department of Physics, Visvesvaraya National Institute of Technology, Nagpur 440010, India}
\date{\today}

\begin{abstract}
We study exact quantum recurrences in the double kicked top (DKT), a driven spin model that extends the quantum kicked top (QKT) by introducing an additional time-reversal symmetry-breaking kick. Reformulating its dynamics in terms of effective parameters $k_r$ and $k_\theta$, we analytically show exact periodicity of the Floquet operator for $k_r = j\pi/2$ and $k_r = j\pi/4$ with distinct periods for integer and half-odd integer $j$. These exact recurrences were found to be independent of $k_\theta$. The long-time-averaged entanglement and fidelity rate function show dynamical quantum phase transition (DQPT) for $k_r = j\pi/2$ at time-reversal symmetric cases $k_\theta = \pm k_r$. In the other time-reversal symmetric case $k_\theta = 0$, the DQPT exists only for a half-odd integer $j$. Using level statistics, a smooth transition is observed from integrable to non-integrable nature as $k_r$ is changed away from $j\pi/2$. Our work demonstrates that regular and chaotic regimes can be controlled for any system size by tuning $k_r$ and $k_\theta$, making the DKT a useful platform for quantum control and information processing applications.
\end{abstract}

\maketitle
\section{Introduction}
Classical recurrence is the phenomenon in which the phase space point returns arbitrarily close to the initial point for any finite-volume, conservative system~\citep{barreira2006poincare,anishchenko2013poincare,saussol2009introduction}. Its quantum analogue arises naturally from the unitary evolution \citep{bocchieri1957quantum,peres1982recurrence,schulman1978note,fishman1982chaos}. These recurrences play an important role in the study of quantum chaos. While classical chaos is hypersensitive to initial conditions, quantum dynamics remain linear and unitary. Thus, it is important to study the spectral properties~\citep{wigner1951statistical, dyson1962statistical,berry1977level,bohigas1984characterization,korenblit2012arbitrary} of the unitary operator of the corresponding dynamical system and recurrence patterns vital~\citep{peres1984stability,peres1984ergodicity,feingold1984ergodicity}.

The QKT is a simple yet rich model that exhibits a smooth transition from regular to chaotic dynamics~\citep{haake1987classical}. The existence of the classical limit allows us to not only study classical-quantum correspondence. But also serves as a testbed for studying quantum information theory due to its connection with the quantum many-body dynamics. Its experimental realizations involve cold atoms~\citep{chaudhary_quantum_signatures}, nuclear magnetic resonance (NMR) platforms~\citep{krithika2023nmr,krithika2022quantum}, and superconducting qubits~\citep{neil2016ergodic}. They offer pathways to faster computation~\citep{monroe2021programmable}, enhanced metrological precision~\citep{pezze2018quantum}, and efficient quantum thermal machines~\citep{zhang2008four}. Its generalizations to systems with long-range interactions involve statistical mechanics~\citep{campa2009statistical, kastner2011diverging}, quantum many-body physics~\citep{fey2016ising, britton2012engineered}, cosmology~\citep{peebles1980large, bouchet2010thermodynamics}, atomic and nuclear physics~\citep{bohr1998nuclear, brink1993semiclassical}, and plasma physics~\citep{nicholson1983introduction}.

Although early studies discuss the arbitrary closeness of the phase space point to its initial position, the recurrence time remains arbitrarily large. Recent studies find periods for exact recurrences by analyzing Floquet spectrum~\citep{sankaranarayanan2003recurrence,amit2024,anand2025quantum,sharma2024exactly,sharma2024signatures}. It was shown in the QKT system~\citep{amit2024} that state-independent exact recurrences generally occur at large chaoticity and do not have a classical analog. These studies investigated the relationship between the chaoticity parameter proportional to a rational multiple of $j \pi$ and the existence of quantum recurrences. In addition, the relationship between QKT periodicities and the quantum resonances in the kicked rotor has also been established~\citep{amit2024}.

To study the effects of broken time-reversal symmetry on quantum chaos, the DKT was considered by introducing a second instantaneous kick per driving period~\citep{haake1987classical}. A recent study~\citep{purohit2025double} has revealed that the DKT exhibits two distinct features of chaos, modified phase-space structures, and entanglement dynamics being highly sensitive to the relative strengths of the two kicks. This introduction of a second kick ($k' \neq 0$) allows much finer control over entanglement evolution for small system size—crucial for quantum information processing. The study~\citep{purohit2025double} leads us to ask whether the DKT exhibits exact quantum recurrences, and if so, how these recurrences depend on the time-reversal symmetry-breaking parameter $k_\theta$. It remains unclear whether broken time-reversal symmetry brings out local dynamical changes or global structural changes.

In this work, we analytically and numerically investigate the periodicity of the DKT Floquet operator across a range of kicking strengths. We show that the operator becomes exactly periodic at special values of the effective kicking parameter $k_r = j\pi/2$, regardless of $k_\theta$. By distinguishing between integer and half-odd-integer spin $j$, we classify the corresponding periodic regimes in terms of the powers of the Floquet operator. We further explore these behaviors using Husimi distributions and von Neumann entropy.

Although the quantum recurrences occur across $k_\theta$, the dynamics remain distinct. To see this, we used fidelity decay as it gives the probability that the time-evolved state returns to its initial state~\citep{gorin2006dynamics,jalabert2001environment,andraschko2014dynamical}. The rate function associated with fidelity is similar to the free energy in thermodynamics. Thus, the fidelity together with its rate function \citep{prosen2003theory,gorin2006dynamics,jalabert2001environment,andraschko2014dynamical,anand2025quantum} allows us to study effects of broken time-reversal symmetry on the stability and DQPTs \citep{Yang,berdanier2017floquet,Naji,Jafari}.

In Sec.~\ref{sec:background}, we introduce the DKT and its dynamics in terms of effective kicking parameters. In Sec.~\ref{sec:jpiby2}, we examine the Floquet operator’s periodicity for $k_r = j\pi/2$ with  integer and for half-odd-integer spin $j$. In Sec.~\ref{sec:jpiby4}, we extend the analysis to $k_r = j\pi/4$. The role of time-reversal symmetry and its breaking is discussed in Sec.~\ref{sec:timereversal}, focusing on entanglement dynamics and fidelity rate function. In Sec.~\ref{sec:near_periodicity}, we study the entanglement dynamics by perturbing the DKT near exact recurrences. Finally, in Sec.~\ref{sec:rc}, we summarize the results and discuss their implications.

\section{Background}\label{sec:background}
The DKT introduces an extra kick of strength \(k'\) immediately after the first kick of strength \(k\) to break the time-reversal symmetry \citep{haake1987classical}. It is governed by the following Floquet operator:
\begin{equation}\label{U}
    \mathcal{U} = \exp\left(- i \frac{k'}{2j}J_x^2\right)\exp\left(- i \frac{k}{2j}J_z^2\right)\exp\left(- i \frac{\pi}{2} J_y\right).
\end{equation}
It was shown that if we transform kick strengths $(k, k')$ to the following kick parameters~\citep{purohit2025double}:
\begin{equation}\label{kkpTransform}
    k_{r} = \frac{k + k'}{2}\;\; \text{ and} \quad k_{\theta} = \frac{k - k'}{2}.
\end{equation}
Then, the effective kick strength $k_r$ gives QKT equivalent dynamics, and $k_\theta$ is responsible for breaking time-reversal symmetry. The case \(k_\theta = \pm k_r\) recovers the standard QKT, and the other time-reversal symmetric case \(k_\theta = 0\) is unique to the DKT.

It was demonstrated that the diagonalization of the Floquet operator \(\mathcal{U}\) for systems with 2 to 4 qubits allows complete characterization of entanglement dynamics~\citep{purohit2025double}. These small systems have revealed periodic behavior of the entanglement dynamics at \(k_r = j\pi/2\), regardless of \(k_\theta\). The exact recurrences in small systems motivate us to investigate whether such periodicities persist in larger systems. This, in turn, offers greater control over entanglement dynamics for the system of any size.

\section{Transformed kick strength \texorpdfstring{$k_r = j\pi/2$}{}}\label{sec:jpiby2}
In this section, we analytically demonstrate that the Floquet operator is periodic for both integer and half-odd integer values of $j$. Additionally, we show that the period remains independent of $k_\theta$. For $k_\theta = k_r$, the DKT reduces to the standard kicked top with $k = 2k_r$. Therefore, for this case, we recover the results of the standard QKT with \(k = j\pi\)~\citep{amit2024}.

\subsection{\texorpdfstring{$\mathbf{Integer } j$}{}}\label{subsec:even-jpiby2}
For integer values of $j$, we express the Floquet operator using Pauli matrices for $a$-th qubit denoted as $\sigma_x^{(a)}$, $\sigma_z^{(a)}$, and $\sigma_y^{(a)}$~\citep{amit2024}. In this representation, the non-linear and precession terms can be rewritten as follows:
\begin{align}
 \exp\left[- i \left(\frac{k_r - k_\theta}{2j}\right)J_x^2\right] &= \exp\left[- i \left(\frac{k_r - k_\theta}{8j}\right) \left(\sum_{a=1}^{2j} \sigma_x^{(a)}\right)^2 \right], \notag \\
 \exp\left[- i \left(\frac{k_r + k_\theta}{2j} \right)J_z^2\right] &=  \exp\left[- i \left(\frac{k_r + k_\theta}{8j}\right) \left(\sum_{a=1}^{2j} \sigma_z^{(a)} \right)^2 \right], \notag \\
 \text{and }\;\exp\left[- i \frac{\pi}{2} J_y\right] &= \exp\left[- i \frac{\pi}{4} \sum_{a=1}^{2j} \sigma_y^{(a)}\right]. 
\end{align}
For even-$2j$, utilizing the properties of Pauli matrices and setting $k_r = j\pi/2$, we can show that $\mathcal{U}^2$ can be written as follows (see supplementary material \citep{supplementary2025}):
\begin{align}
 \mathcal{U}^2 =& \; \frac{e^{-i\frac{\pi}{4}2j}}{4} \left[ C_1 + C_2 \cos\left(\frac{k_\theta}{2j}\right) + C_3\sin\left(\frac{k_\theta}{2j}\right) \right] \\
    & \times \left[ C_1 + C_2 \cos\left(\frac{k_\theta}{2j}\right) - C_3\sin\left(\frac{k_\theta}{2j}\right) \right] \left(\gamma^{\otimes 2j}\right)^2, \notag
\end{align}
where,
\begin{align}
 C_1 =& \left[\mathds{I}^{\otimes 2j} + (i\sigma_y)^{\otimes 2j}\right] \cos\left(\frac{\pi}{4}\right) \notag \\
 &\quad + i \left[(i\sigma_z)^{\otimes 2j} + (i\sigma_x)^{\otimes 2j}\right] \sin\left(\frac{\pi}{4}\right), \notag \\
 C_2 =& \;\mathds{I}^{\otimes 2j} - (i\sigma_y)^{\otimes 2j}, \quad
 C_3 = i \left[(i\sigma_z)^{\otimes 2j} - (i\sigma_x)^{\otimes 2j}\right] \text{ and} \notag \\
    \gamma =& \; e^{-i\frac{\pi}{4}\sigma_y}.
\end{align}
Using identities $C_1 C_2 + C_2 C_1 = 0$, $C_3 C_1 - C_1 C_3 = 0$, $C_3 C_2 - C_2 C_3 = 0$ and $C_3^2 = - C_2^2$, we get $\mathcal{U}^2$ as follows (see supplementary material~\citep{supplementary2025}):
\begin{align}\label{eq:jpiby2_even_u2}
 \mathcal{U}^2 = \; \frac{e^{-i\frac{\pi}{4}2j}}{4} \left( C_1^2 + C_2^2 \right) \cdot \left(\gamma^{\otimes 2j}\right)^2. 
\end{align}
It can be seen that $\mathcal{U}^2$ is independent of $k_\theta$. This makes all even powers of $\mathcal{U}$ identical with the corresponding Floquet operator of QKT~\citep{amit2024}. Thus, taking the fourth power of the Floquet operator, we get 
\begin{align}
 \mathcal{U}^4 = - e^{i\frac{\pi}{2}2j} {(i\sigma_y)}^{\otimes 2j} \implies \mathcal{U}^8 = \mathds{I}.
\end{align}

To compare classical and quantum dynamics, we use the Husimi quasi-probability distribution defined as follows~\citep{amit2024,agarwal1981relation}:
\begin{align}
 Q_\rho (\theta, \phi) = \langle \theta, \phi | \rho | \theta, \phi \rangle,
\end{align}
where $|\theta, \phi \rangle$ is the spin-coherent state associated with SU(2) dynamical symmetry, and $\rho = \rho(n)$ is the density operator evolved from an initial coherent state after $n$ steps. The distribution satisfies the normalization condition:
\begin{align}
 \frac{2j+1}{4\pi} \int_{S^2} Q_\rho (\theta, \phi) \; \sin \theta \, d\theta \, d\phi = 1.
\end{align}
\begin{figure}[!ht]
    \includegraphics[width=\linewidth]{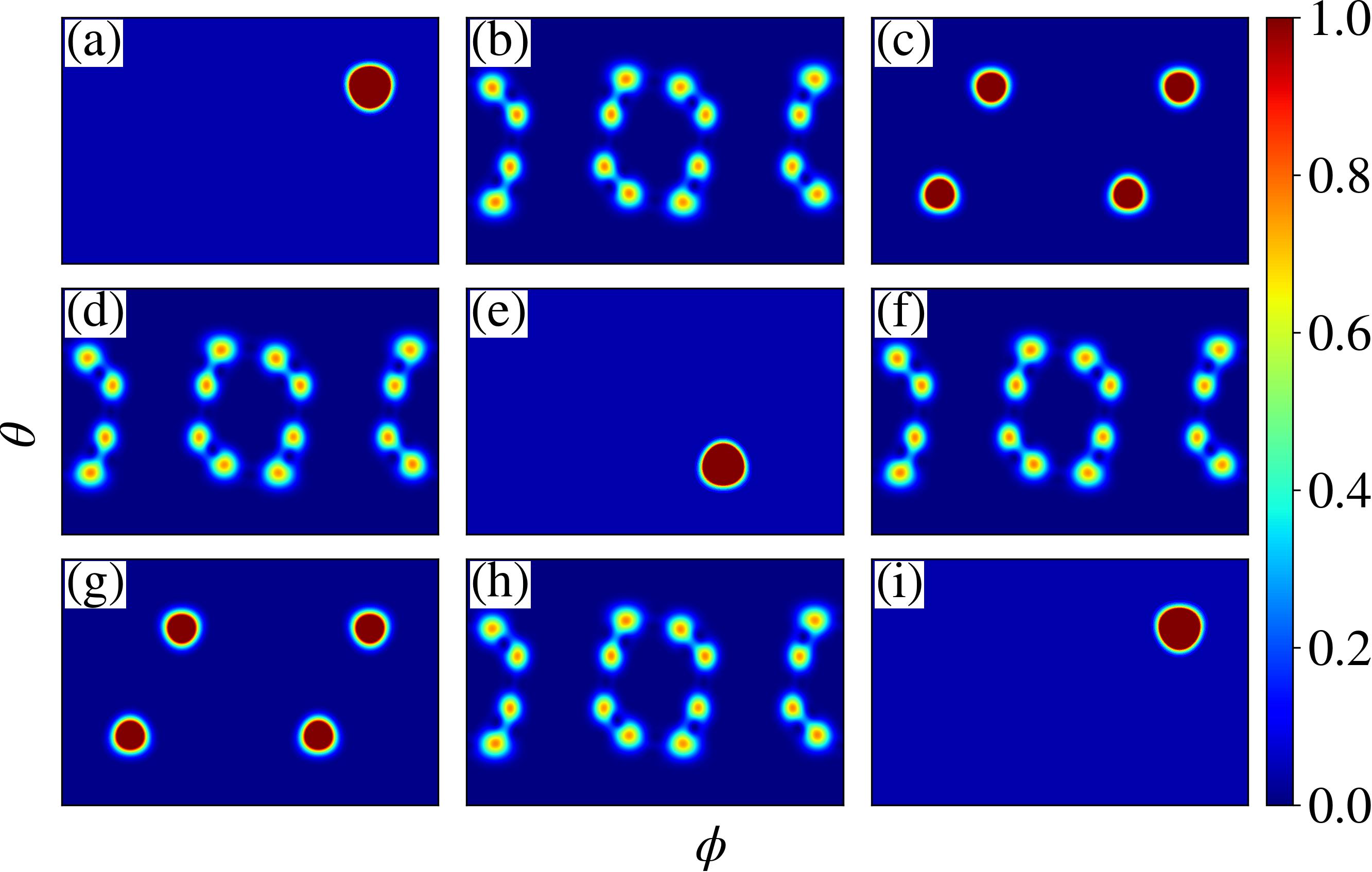}
    \caption{Husimi function of the time-evolved initial state $|\theta_0 = 2.25,\, \phi_0 = 2.0\rangle$ over eight kicks. Here, $k_r = j\pi/2$, $k_\theta = 0$, and $j = 76$. Panels (a)–(i) correspond to $n = 0$ through $n = 8$, respectively.}
    \label{husimi_kt_0}
\end{figure}
\begin{figure}[!ht]
    \includegraphics[width=\linewidth]{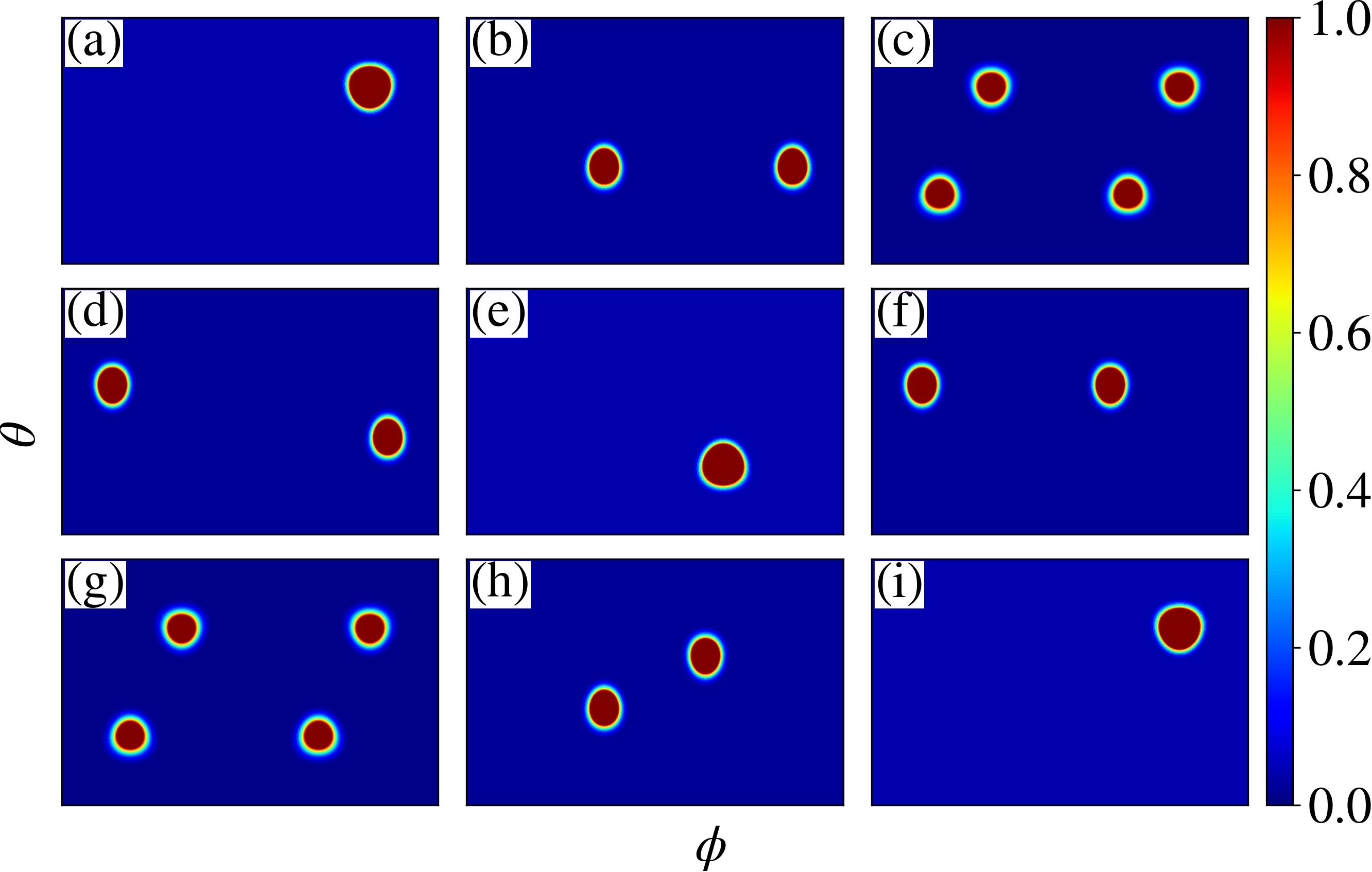}
    \caption{Husimi function of the time-evolved initial state $|\theta_0 = 2.25,\, \phi_0 = 2.0\rangle$ over eight kicks. Here, $k_r = j\pi/2$, $k_\theta = k_r$, and $j = 76$. Panels (a)–(i) correspond to $n = 0$ through $n = 8$, respectively.}
    \label{husimi_kt_kr}
\end{figure}

The Husimi distributions were computed for the two cases, $k_\theta = 0$ and $k_\theta = k_r$. They show distribution identical for even values of $n$ in both cases (see Figs.~\ref{husimi_kt_0} and \ref{husimi_kt_kr}). More importantly, the distribution returns to its initial configuration at $n = 8$.
 
Even though the periodic behavior (i.e., period) is independent of $k_\theta$, the dynamics themselves depend on $k_\theta$. As a result, the long-time-averaged quantum correlations depend on $k_\theta$. To quantify the effects of broken time-reversal symmetry, we define the long-time-averaged von Neumann entropy as follows~\citep{neil2016ergodic,Zarum1998}:
\begin{align}
    \langle S_{(\theta_0, \phi_0)} \rangle &= \lim_{N \to \infty} \frac{1}{N} \sum_{n = 0}^{N-1} S_{(\theta_0, \phi_0)}(n) \; \text{ and} \\
 S_{(\theta_0, \phi_0)}(n) &= - \text{Tr}[\rho_1(n) \log_2 \rho_1(n)],
\end{align}
where $\rho_1(n)$ is the single-qubit reduced density matrix corresponding to the evolved state $|\theta_0, \phi_0\rangle$ after $n$ time steps.

\begin{figure}[!ht]
    \includegraphics[width=\linewidth]{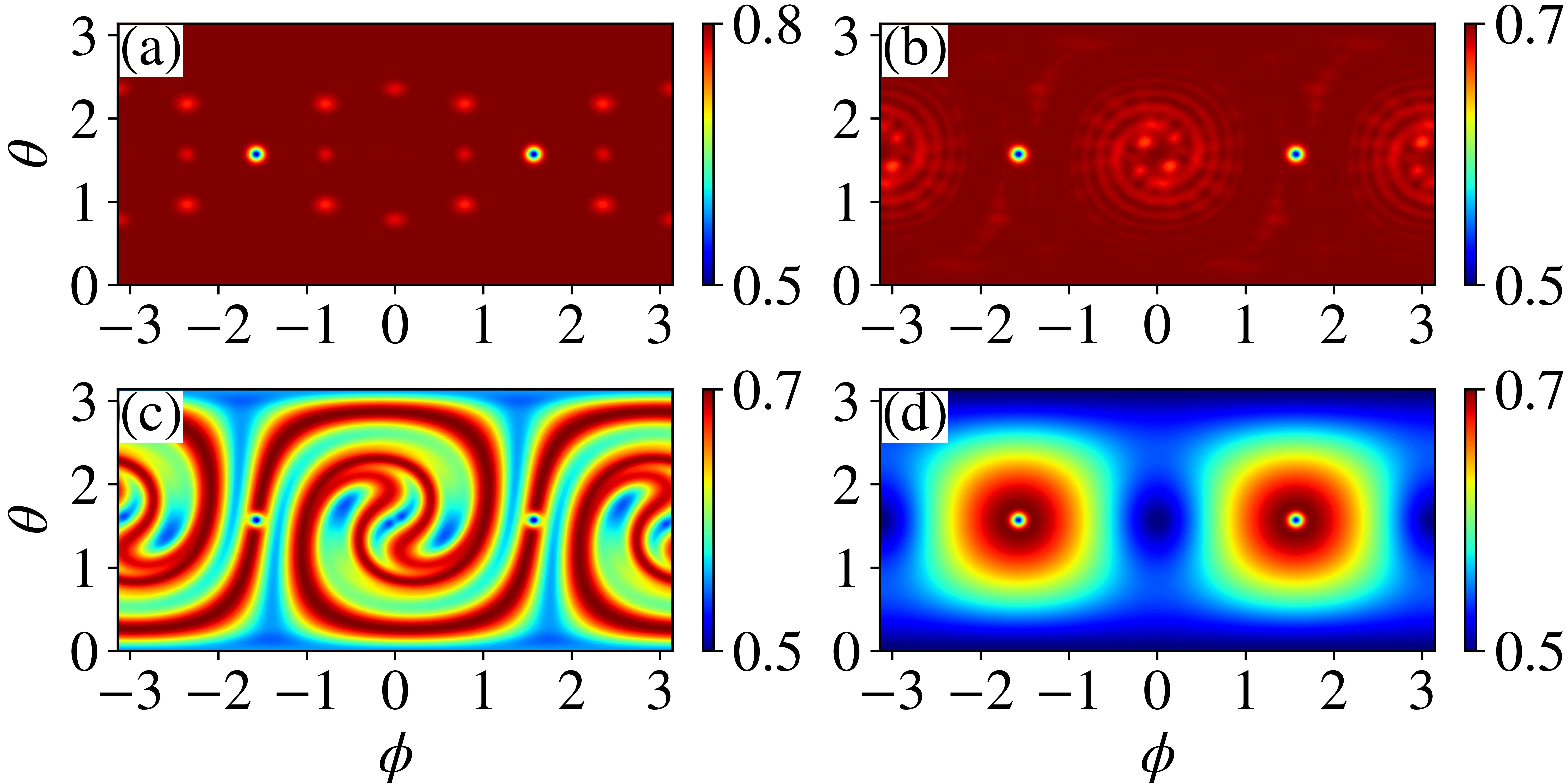}
    \caption{Long-time-averaged von Neumann entropy for the single-particle reduced density matrix $\rho_1(n)$, with total spin $j = 76$. We evolve 40,000 initial spin-coherent states $|\theta_0, \phi_0\rangle$ over $n = 1000$ steps. Here, $k_r = j\pi/2$ and (a) $k_\theta = 0$, (b) $k_\theta = 0.75 j\pi/2$, (c) $k_\theta = 0.95 j\pi/2$, (d) $k_\theta = j\pi/2$.}
    \label{fig76_jpiby2}
\end{figure}

\begin{figure}[!ht]
    \includegraphics[width=\linewidth]{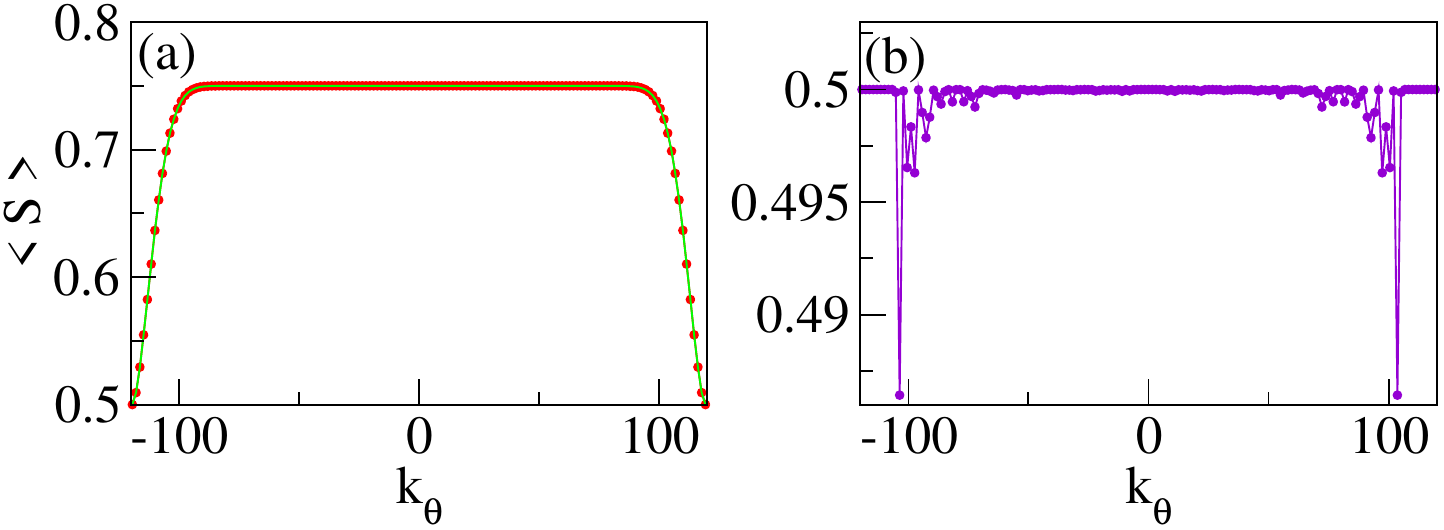}
    \caption{Long-time-averaged von Neumann entropy for the initial spin-coherent states (a) $|\theta_0 = 0, \phi_0 = 0\rangle$ and (b) $|\theta_0 = \pi/2, \phi_0 = \pm \pi/2\rangle$, with $k_r = j\pi/2$ and total spin $j = 76$, evolved for $n = 1000$ time steps.}
    \label{entropy-even}
\end{figure}

Our computations show that states $|\theta_0 = \pi/2, \phi_0 = \pm \pi/2 \rangle$ largely remain unaffected by the broken time-reversal symmetry (see Figs.~\ref{fig76_jpiby2} and \ref{entropy-even}). On perturbing $k_\theta$ away from zero, the dynamics remain qualitatively similar until $k_\theta$ approaches $k_r = j \pi/2$ (see Fig.~\ref{entropy-even}). However, breaking time-reversal symmetry at $k_\theta = \pm k_r$ leads to significant changes in the entanglement of most of the states except $|\theta_0 = \pi/2, \phi_0 = \pm \pi/2 \rangle$ (see Fig.~\ref{fig76_jpiby2}). As $k_\theta$ approaches $j\pi/2$, vortex-like structures emerge around states corresponding to the trivial fixed points (see Figs.~\ref{fig76_jpiby2}(b–c)). This shows that the effect of broken time-reversal symmetry on the entanglement dynamics depends on the initial state.

We now explain why breaking time-reversal symmetry for cases $k_\theta = 0$ and $k_\theta = k_r$ shows different results. Here, we assume a large system size. For the case $k_\theta = k_r$, the time-reversal symmetry is broken by adding a perturbation $k'$ to the standard QKT. This perturbation introduces vortices as illustrated in Figs.~\ref{fig76_jpiby2}(d) $\to$ \ref{fig76_jpiby2}(c). Here, the action of $k_\theta$ creating vortices around the trivial fixed points is similar to the one observed in Figs.~(24) and (25) of Ref.~\citep{purohit2025double}. For the case $k_\theta = 0$, broken time-reversal symmetry is not equivalent to adding a perturbation as $k\approx k'$. As a result, there are no noteworthy changes in the entanglement dynamics near $k_\theta = 0$. 

\subsection{\texorpdfstring{\textbf{Half-odd integer } $j$}{}}\label{subsec:odd-jpiby2}
We begin by showing the third power of the Floquet operator $\mathcal{U}$ is independent of $k_\theta$, and then determine its periodicity (see supplementary material~\citep{supplementary2025}). Using identities $\gamma \sigma_z = \sigma_x \gamma$, $\sigma_z \gamma = -\gamma \sigma_x$~\citep{amit2024}, and by taking $\gamma$ to the right, we get:
\begin{align}
 \mathcal{U}^2 &= \exp\left(- i \frac{k'}{2j}J_x^2\right) \exp\left(- i \frac{k_r}{j}J_z^2\right)  \exp\left(- i \frac{k}{2j}J_x^2\right) \left(\gamma^{\otimes 2j}\right)^2 \notag \\ 
    &= \exp\left(- i \frac{k'}{2j}J_x^2\right) \exp\left(- i \frac{\pi}{2}J_z^2\right)  \exp\left(- i \frac{k}{2j}J_x^2\right) \left(\gamma^{\otimes 2j}\right)^2.
\end{align}
Unlike the case of even-$2j$, here, we do not get the simplified form obtained in Eq.~\eqref{eq:jpiby2_even_u2}. Hence, continuing along the similar lines of Sec.~\ref{subsec:even-jpiby2}, we get
\begin{align}
 \mathcal{U}^3 = e^{- i \frac{k'}{2j}J_x^2} e^{- i \frac{k_r}{j}J_z^2} e^{- i \frac{k_r}{j}J_x^2} e^{- i \frac{k}{2j}J_z^2} \left(\gamma^{\otimes 2j}\right)^3.
\end{align}
Proceeding further by setting $k' = k_r - k_\theta$ and $k_r = j\pi/2$, we get
\begin{align}
 \mathcal{U}^3 = e^{- i \left(\frac{\pi}{4} - \frac{k_\theta}{2j}\right) J_x^2} e^{- i \frac{\pi}{2}J_z^2} e^{- i \frac{\pi}{2}J_x^2} e^{- i \left(\frac{\pi}{4} + \frac{k_\theta}{2j}\right) J_z^2} \left(\gamma^{\otimes 2j}\right)^3.
\end{align}
We provide detailed derivations in the supplementary material~\citep{supplementary2025}. Using the operator identities given by
\begin{align}
 \exp\left(i\frac{k_\theta}{4j}\sigma_x\right) \exp\left(-i\frac{\pi}{4}\sigma_z \right) &= \exp\left(-i\frac{\pi}{4}\sigma_z\right) \exp\left(-i\frac{k_\theta}{4j}\sigma_y\right), \notag \\
 \exp\left(-i\frac{\pi}{4}\sigma_x\right) \exp\left(-i\frac{k_\theta}{4j}\sigma_z\right) &= \exp\left(i\frac{k_\theta}{4j}\sigma_y\right) \exp\left(-i\frac{\pi}{4}\sigma_x\right), 
\end{align}
we find that:
\begin{align}
 \exp\left(i \frac{k_\theta}{2j}J_x^2\right) \exp\left(- i \frac{\pi}{2}J_z^2\right) &= \exp\left(- i \frac{\pi}{2}J_z^2\right) \exp\left(i \frac{k_\theta}{2j}J_y^2\right), \notag \\
 \exp\left(- i \frac{\pi}{2}J_x^2\right) \exp\left(-i \frac{k_\theta}{2j}J_z^2\right) &= \exp\left(-i \frac{k_\theta}{2j}J_y^2\right) \exp\left(- i \frac{\pi}{2}J_x^2\right).
\end{align}
Using these relations, we get
\begin{align}
 \mathcal{U}^3 = e^{- i \frac{\pi}{4} J_x^2} e^{- i \frac{\pi}{2}J_z^2} e^{- i \frac{\pi}{2}J_x^2} e^{- i \frac{\pi}{4} J_z^2} \left(\gamma^{\otimes 2j}\right)^3.
\end{align}
Note that the above operator is independent of $k_\theta$. Continuing further, we obtain the operator $\mathcal{U}^6$ as follows:
\begin{align}
    \mathcal{U}^6 =& \; \exp\left(- i \frac{\pi}{4}J_x^2\right) {\left[\exp\left(- i \frac{\pi}{2}J_z^2\right) \exp\left(- i \frac{\pi}{2}J_x^2\right)\right]}^3 \notag \\
    &\quad \times \exp\left( i \frac{\pi}{4}J_x^2\right) \cdot {\left(\gamma^{\otimes 2j}\right)}^6.
\end{align}
Then, using property given by (see Eq.~(B31) of Ref.~\citep{amit2024})
\begin{align}
    {\left[\exp\left( - i \frac{\pi}{2}J_z^2\right) \exp\left( - i \frac{\pi}{2}J_x^2\right)\right]}^3 = - \mathds{I}^{\otimes 2j},
\end{align}
we get 
\begin{align}
    \mathcal{U}^6 = - {(-1)}^{2j} \; {(i\sigma_y)}^{\otimes 2j} \implies \mathcal{U}^{12} = {(-1)}^{2j} \; \mathds{I}^{\otimes 2j}.
\end{align}
Thus, the Floquet operator $\mathcal{U}$ is periodic with a period of 12~\citep{amit2024} and is independent of $k_\theta$.

\begin{figure}[!ht]
    \includegraphics[width=\linewidth]{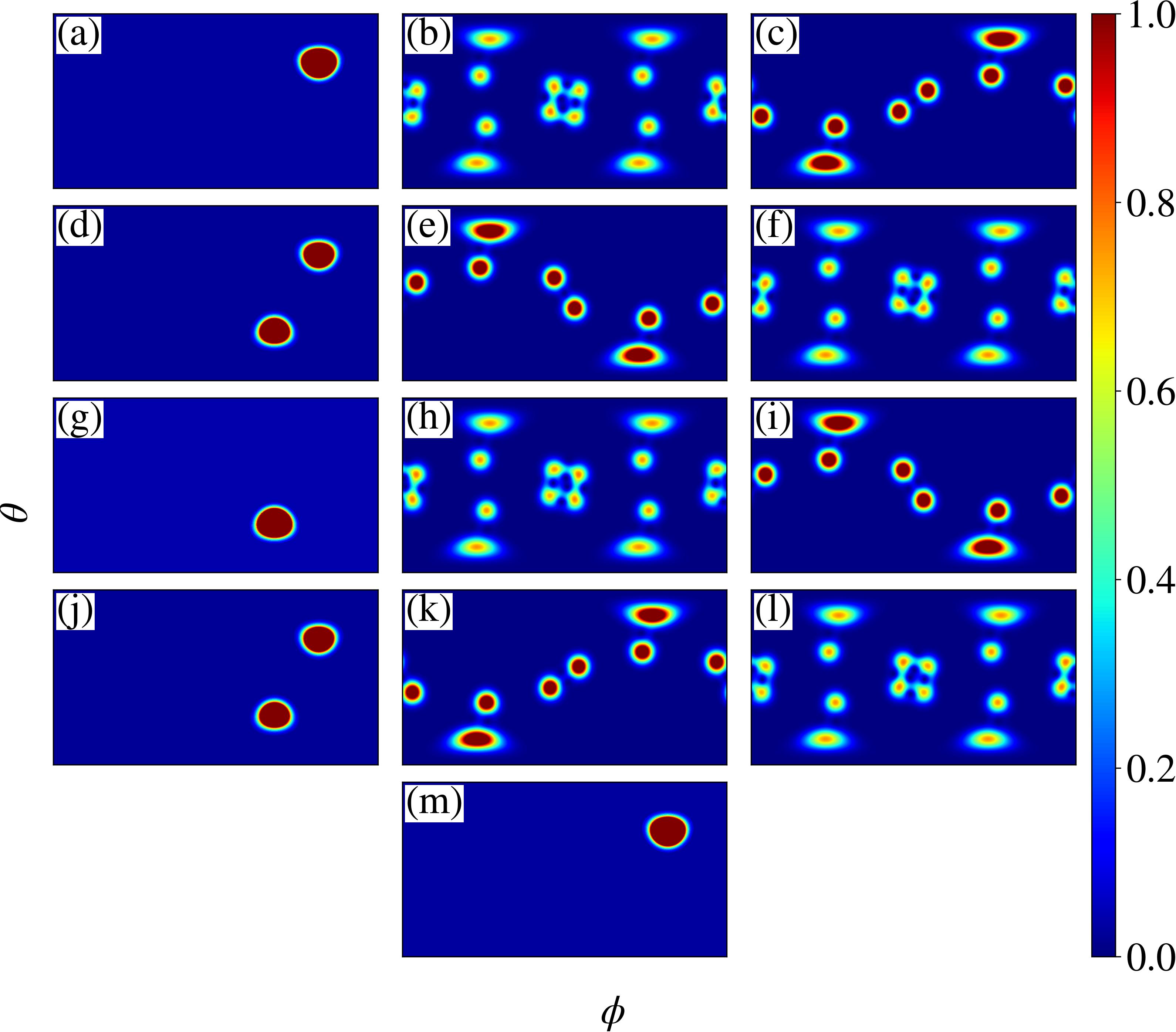}
    \caption{Husimi function of the time-evolved initial state $|\theta_0 = 2.25,\, \phi_0 = 2.0\rangle$ over eight kicks. Here, $k_r = j\pi/2$, $k_\theta = 0$, and $j = 75.5$. Panels (a)–(i) correspond to $n = 0$ through $n = 12$, respectively.}\label{jpiby2_odd1}
\end{figure}

\begin{figure}[!ht]
    \includegraphics[width=\linewidth]{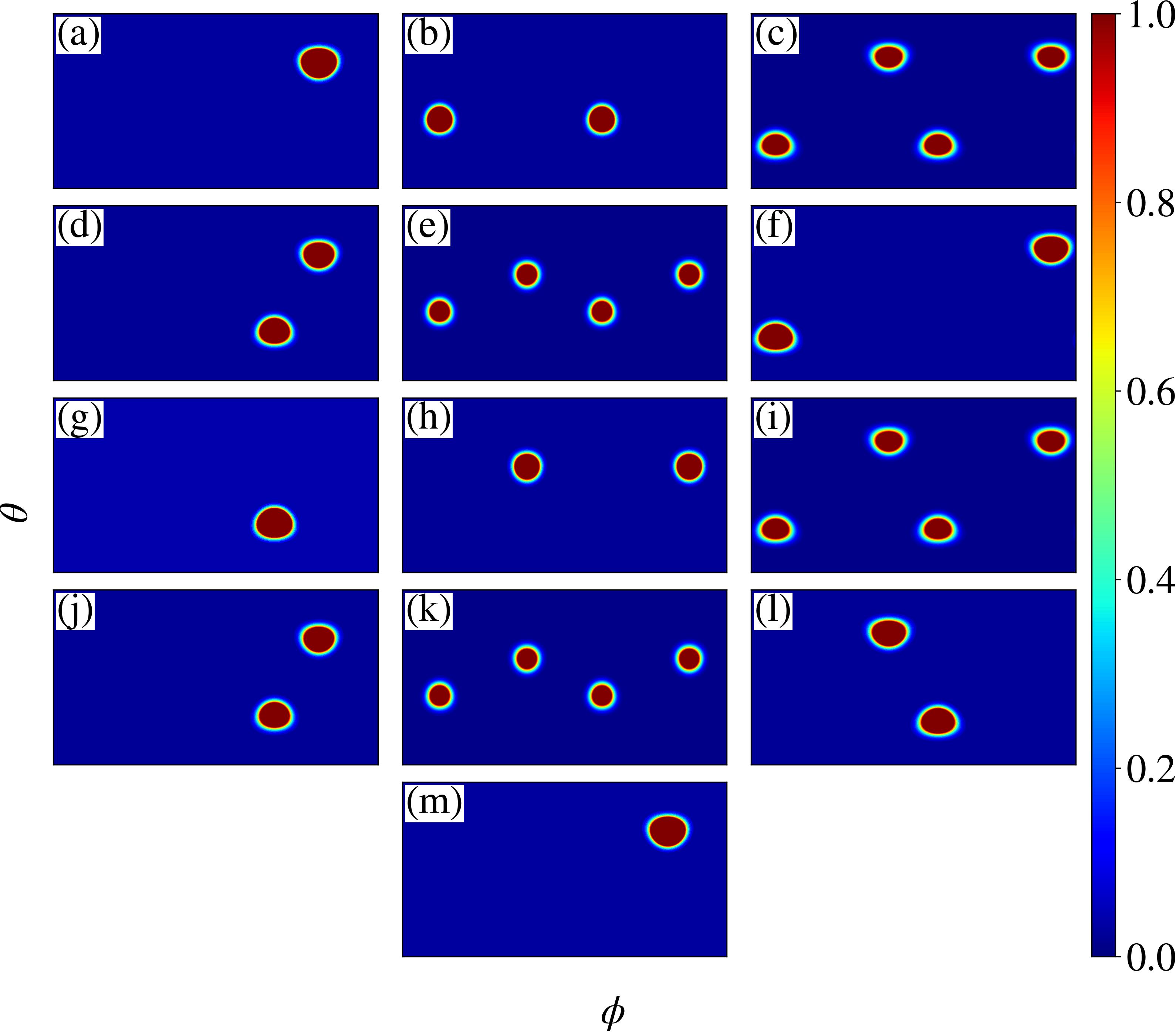}
    \caption{Husimi function of the time-evolved initial state $|\theta_0 = 2.25,\, \phi_0 = 2.0\rangle$ over eight kicks. Here, $k_r = j\pi/2$, $k_\theta = k_r$, and $j = 75.5$. Panels (a)–(i) correspond to $n = 0$ through $n = 12$, respectively.}\label{jpiby2_odd2}
\end{figure}

Now, we plot the Husimi distribution for both $k_\theta = 0$ and $k_\theta = k_r$ as it resembles classical probability distribution (see Figs.~\ref{jpiby2_odd1} and \ref{jpiby2_odd2}). Computations show that the distributions for both cases are identical at time steps $n = 0$, $n = 3$, $n = 6$, $n = 9$, and $n = 12$. At $n = 12$, the distribution returns to its initial configuration.

\begin{figure}[!ht]
    \includegraphics[width=\linewidth]{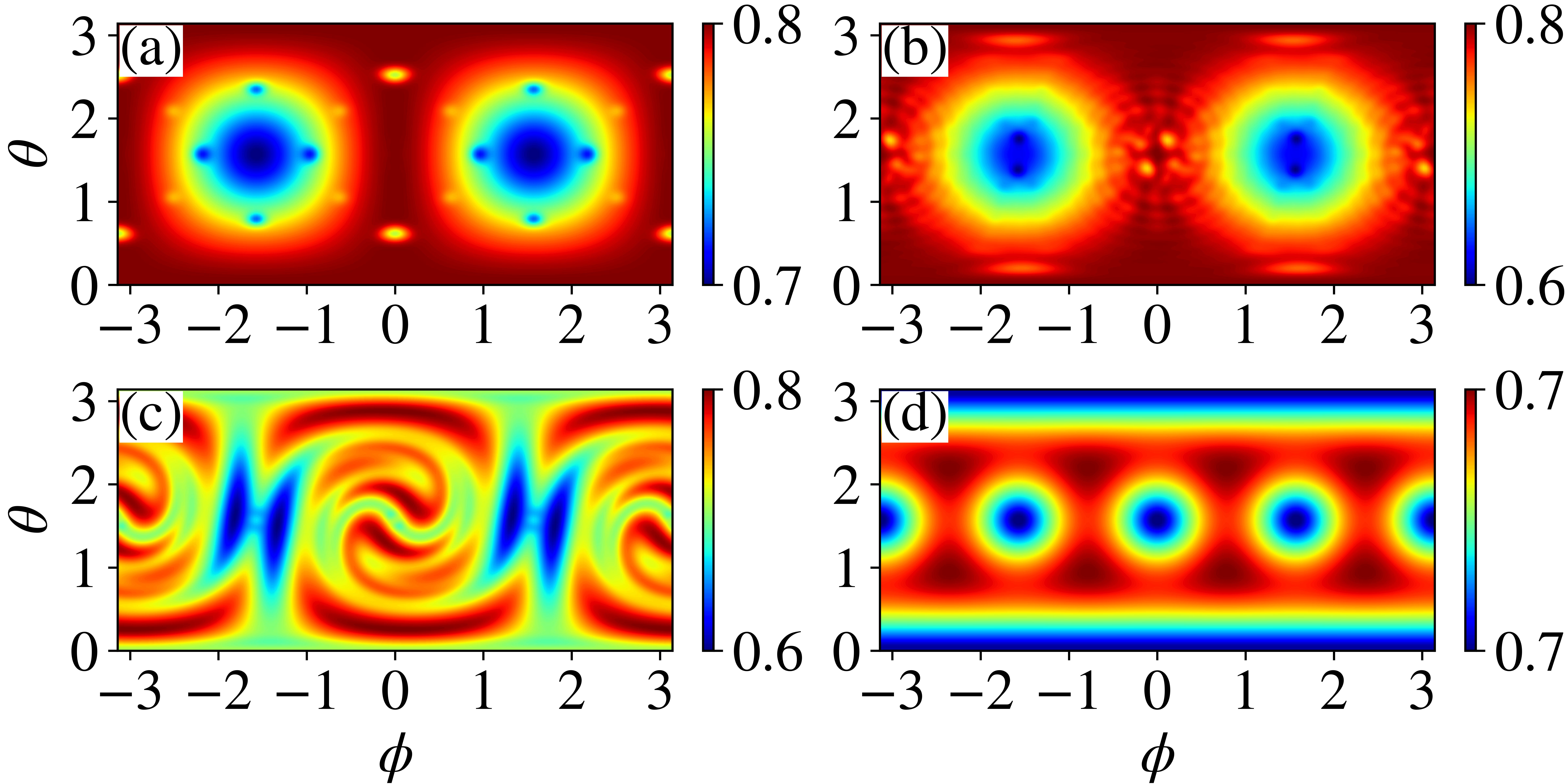}
    \caption{Long-time-averaged von Neumann entropy for the single-particle reduced density matrix $\rho_1(n)$, with total spin $j = 75.5$. We evolve 40,000 initial spin-coherent states $|\theta_0, \phi_0\rangle$ over $n = 1000$ steps. Here, $k_r = j\pi/2$ and (a) $k_\theta = 0$, (b) $k_\theta = 0.75 j\pi/2$, (c) $k_\theta = 0.95 j\pi/2$, (d) $k_\theta = j\pi/2$.}
    \label{fig75.5_jpiby2}
\end{figure}

\begin{figure}[!ht]
    \includegraphics[width=\linewidth]{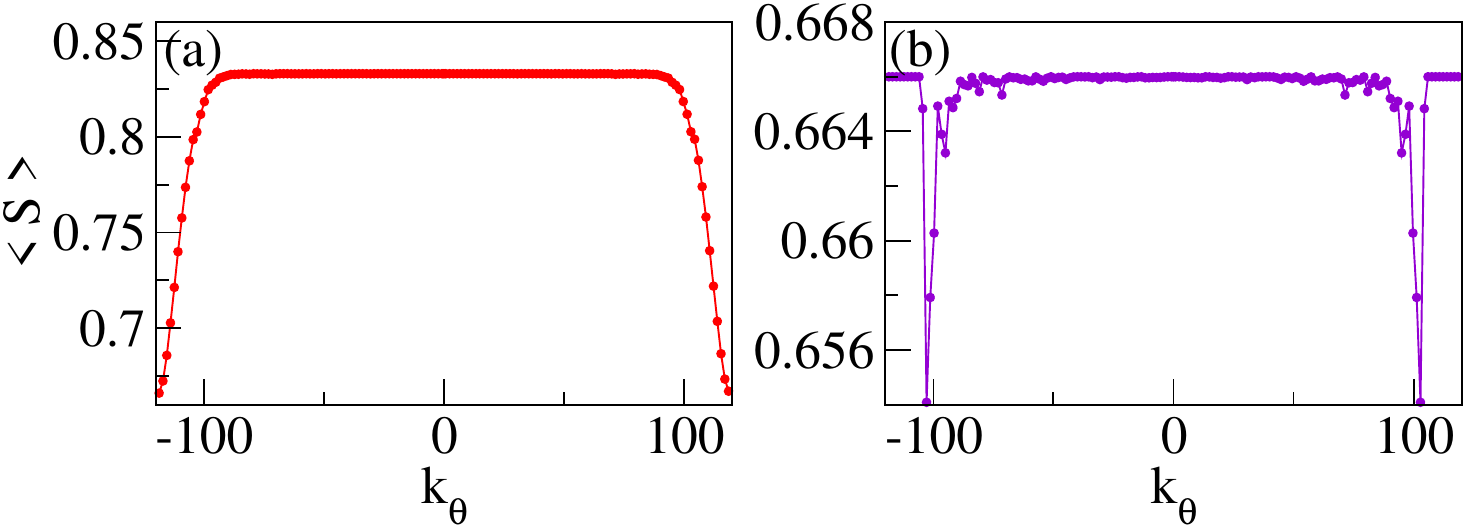}
    \caption{Long-time-averaged von Neumann entropy computed over $n = 1000$ steps for the initial spin-coherent states (a) $|\theta_0 = 0, \phi_0 = 0\rangle$ and (b) $|\theta_0 = \pi/2, \phi_0 = \pm \pi/2\rangle$, with $k_r = j\pi/2$ and total spin $j = 75.5$.}
    \label{entropy-odd}
\end{figure}

The qualitative features of the long-time-averaged von Neumann entropy landscape for half-odd integer $j$ are similar to its corresponding integer $j$ case (see Figs.~\ref{fig75.5_jpiby2}). These features include responses of $k_\theta = 0$ and $k_\theta = \pm k_r$ to perturbations and vortex-like effects (see Figs.~\ref{fig75.5_jpiby2} and \ref{entropy-odd}).

\section{Transformed kick strength \texorpdfstring{$k_r = j\pi/4$}{}}\label{sec:jpiby4}
We examine the periodicity of $\mathcal{U}$ for the case \(k_r = j\pi/4\). In the case of integer $j$, we show that the period of the Floquet operator is independent of \(k_\theta\). However, for a half-odd integer $j$, we do not observe periodic behavior. As mentioned earlier, for \(k_\theta = k_r\), the DKT reduces to the standard QKT with \(k = j\pi/2\). Although this case has been studied numerically for the standard QKT~\citep{amit2024}, the analytical proof for integer $j$ was not given, and the reason for the absence of periodicity for half-odd integer $j$ was also not provided. In this section, we provide analytical calculations for these cases of the DKT. As a consequence, we recover the results for the standard QKT.

\subsection{\texorpdfstring{$\textbf{Integer } j$}{}}\label{subsec:even-jpiby4}
First, we show that the operator $\mathcal{U}^{12}$ is independent of $k_\theta$ and then calculate the period of this case. Following a similar analysis as in Sec.~\ref{sec:jpiby2}, we obtain the following expression for the third power of the Floquet operator (see supplementary material~\citep{supplementary2025}):
\begin{align}
    \mathcal{U}^3 =& \; \exp\left(- i \frac{k'}{2j}J_x^2\right) \left[\exp\left(- i \frac{\pi}{4}J_z^2\right) \exp\left(- i \frac{\pi}{4}J_x^2\right)\right] \notag \\
    &\quad \times \exp\left(- i \frac{k}{2j}J_z^2\right) {\left(\gamma^{\otimes 2j}\right)}^3.
\end{align}
By repetitive multiplications of this operator and simplifying by swapping ${\left(\gamma^{\otimes 2j}\right)}^3$ to the right by transforming $J_z^2 \to J_x^2$, $J_x^2 \to J_z^2$, we get
\begin{align}
    \mathcal{U}^{12} = \; \exp\left(- i \frac{k'}{2j}J_x^2\right) \mathcal{O} \exp\left(- i \frac{k}{2j}J_x^2\right) \exp\left(i \frac{\pi}{4}J_x^2\right){\left(\gamma^{\otimes 2j}\right)}^{12},
\end{align}
where,
\begin{align}\label{eq:operatorjpiby4}
    \mathcal{O} = {\left[\exp\left(- i \frac{\pi}{4}J_z^2\right) \exp\left(- i \frac{\pi}{4}J_x^2\right)\right]}^6.
\end{align}
Since the operator $\mathcal{O}$ commutes with $J_x^2$ operator (see supplementary material~\citep{supplementary2025}), the above Floquet operator simplifies as follows:
\begin{align}
    \mathcal{U}^{12} =& \; \mathcal{O} \cdot {\left(\gamma^{\otimes 2j}\right)}^{12}.
\end{align}
For even-$2j$, the operator $\mathcal{O}$ can be simplified using properties of the Pauli matrices and the Caley-Hamilton theorem (see supplementary material~\citep{supplementary2025}) as follows:
\begin{align}
    \mathcal{O} = \frac{e^{-i\frac{3\pi}{4}2j}}{2} \left[\mathds{I}^{\otimes 2j} + {(-1)}^j {\sigma_y}^{\otimes 2j} + i {\sigma_z}^{\otimes 2j} + i {\sigma_x}^{\otimes 2j} \right].
\end{align}
On squaring further, we get
\begin{align}
    \mathcal{U}^{24} = - \sigma_y^{\otimes 2j} \implies \mathcal{U}^{48} = \mathds{I}^{\otimes 2j}.
\end{align}
The Floquet operator \(\mathcal{U}\), therefore, has period 48.

\subsection{\texorpdfstring{$\textbf{Half-odd integer } j$}{}}\label{subsec:odd-jpiby4}
In contrast to the integer case, numerical results reported in \citep{amit2024} show that the Floquet operator does not exhibit periodicity for half-odd integer values of \(j\). In this subsection, we provide an analytical explanation for this observation with the DKT extension. The 12th power of the Floquet operator for this case is given by
\begin{align}
    \mathcal{U}^{12} =& \; \exp\left(- i \frac{k'}{2j}J_x^2\right) \mathcal{O} \exp\left(- i \frac{k}{2j}J_x^2\right) \exp\left(i \frac{\pi}{4}J_x^2\right){\left(\gamma^{\otimes 2j}\right)}^{12} \notag \\
    \mathcal{O} =&\; {\left[\exp\left(- i \frac{\pi}{4}J_z^2\right) \exp\left(- i \frac{\pi}{4}J_x^2\right)\right]}^6.
\end{align}
Now, we construct a state $|\psi\rangle_{2j}$ for which
\begin{align}
    \mathcal{O}^{m} |\psi\rangle_{2j} \neq |\psi\rangle_{2j} 
    \quad \text{for any } m \in \mathds{Z}_{>0}.
\end{align}
To do so, we split the odd-$2j$ system into $(j-3/2)$-pairs of qubits and unpaired 3-qubits. Let $(a,b) = (1,2), (3,4), \cdots$ represent the paired qubits and $t_1, t_2, t_3$ represent the unpaired qubits. For each pair $(a,b)$, the two-spin singlet state is defined as follows:
\begin{align}
    |\psi_{ab}^-\rangle = \frac{1}{\sqrt{2}} |01\rangle - \frac{1}{\sqrt{2}} |10\rangle.
\end{align}
It satisfies the property given by
\begin{align}
    &\left(\sigma_i^{(a)} + \sigma_i^{(b)}\right) |\psi_{ab}^-\rangle = 0 & \text{for } i = x, y, z.
\end{align}
Using this two-qubit singlet state, the full odd-$2j$ state can be constructed as follow
\begin{align}
    |\psi\rangle_{2j} = \bigotimes_{(a,b)}^{j-3/2} |\psi_{ab}^-\rangle \otimes |\chi_{t_1, t_2, t_3} \rangle.
\end{align}
By splitting the full angular momentum operator given by
\begin{align}
    J_i = \sum_{(a,b)} J_i^{(a,b)}\otimes \mathds{I}^{\otimes 3} + \mathds{I}^{\otimes (2j-3)} \otimes J_i^{(3)},
\end{align}
the action of $J_i$ on the singlet subspace reduces to the action of $J_i$ on the unpaired three qubits: 
\begin{align}
    J_i |\psi\rangle_{2j} = \mathds{I}^{\otimes (2j-3)} \otimes J_i^{(3)} |\psi\rangle_{2j} \quad \text{for } i = x, y, z.
\end{align}
Here, the angular momentum $J_i^{(3)} = \left(\sigma_i^{t_1} + \sigma_i^{t_2} + \sigma_i^{t_3}\right)/2$ acts only on the state $|\chi_{t_1, t_2, t_3} \rangle$ and the angular momentum $J_i^{(a,b)}$ of the pair $(a,b)$ acts trivially on the state $|\psi_{ab}^-\rangle$. It follows that the action of a polynomial $P_n\left(J_i\right)$ on the state $|\psi\rangle_{2j}$ reduces to the action of its corresponding unpaired three-qubit version on the state $|\chi_{t_1, t_2, t_3} \rangle$: 
\begin{align}
    P_n\left(J_i\right) |\psi\rangle_{2j} = \mathds{I}^{\otimes (2j-3)} \otimes P_n (J_i^{(3)}) |\psi\rangle_{2j}.
\end{align}
Therefore, the action of an operator $\mathcal{O}$ on the state $|\psi\rangle_{2j}$ is given by
\begin{align}
    \mathcal{O} |\psi\rangle_{2j}  = \mathds{I}^{\otimes (2j - 3)} \otimes  {\left(\mathcal{A}^{(3)}\right)}^6 |\psi\rangle_{2j},
\end{align}
where
\begin{align}
    \mathcal{A}^{(3)} = \exp\left(- i \frac{\pi}{4}J_z^{(3)} \cdot J_z^{(3)}\right) \exp\left(- i \frac{\pi}{4}J_x^{(3)} \cdot {J_x^{(3)}}\right).
\end{align}
The eigenvalues of $\mathcal{A}^{(3)}$ are $e^{i \pi/4}$ with multiplicity 4 and $e^{i \pi/4}(\pm \sqrt{7} - 3i)/4$ with multiplicity 2 for each. The last four eigenvalues cannot be written in the form \((-1)^r\) for any \(r \in \mathds{Q}\). This shows that the action of the operator $\mathcal{O}$ is not periodic on the state $|\psi\rangle_{2j}$. 

If there was some integer $t$ satisfying $\mathcal{O}^t = \mathds{I}$ then, it would have satisfied for the state $|\psi\rangle_{2j}$ as well. But this is a contradiction. Hence, $\mathcal{O}^t \neq \mathds{I}$ for all positive integer values of $t$. As a result, the Floquet operator $\mathcal{U}$ does not have periodicity.


\section{Broken time-reversal symmetry}\label{sec:timereversal}
We examine the role of time-reversal symmetry in exact recurrences and its breaking through the fidelity rate function. In earlier sections, we have shown that the long-time-averaged von Neumann entropy landscapes for even-$2j$ and odd-$2j$ exhibit distinct behaviors in both time-reversal-symmetric cases, \(k_\theta = k_r\) and \(k_\theta = 0\) (see Figs.~\ref{fig76_jpiby2} and \ref{fig75.5_jpiby2}). Additionally, the states \( |\theta_0 = 0, \phi_0 = 0\rangle \) and \( |\theta_0 = \pi/2, \phi_0 = - \pi/2\rangle \) are of particular interest. The state \( |\theta_0 = 0, \phi_0 = 0\rangle \) exhibits sharp transition in entanglement only for \(k_\theta = k_r\) (see Figs.~\ref{entropy-even}(a) and \ref{entropy-odd}(a)). The state \( |\theta_0 = \pi/2, \phi_0 = - \pi/2\rangle \), on the other hand, shows formation of vortices just before $k_\theta = k_r$ (see Figs.~\ref{entropy-even}(b) and \ref{entropy-odd}(b)). However, at $k_\theta = k_r$, there are no noteworthy changes. This suggests that the mere breaking of time-reversal symmetry does not necessarily lead to abrupt changes in entanglement dynamics. These qualitatively distinct features motivate us to further investigate using the fidelity rate function, as it gives insights into the DQPTs.

The rate function $R(n)$, derived from the fidelity $\mathcal{Z}(n)$, is defined \citep{Heyl2013dynamical} as follows:
\begin{align}
 \mathcal{Z}(n) =& \; \left|\langle \theta_0, \Phi_0 | {\mathcal{U}(j, k_r, k_\theta )}^n | \theta_0, \Phi_0 \rangle \right|^2 \;\; \mbox{and} \\
 R(n) =& - \frac{1}{2j+1} \ln \mathcal{Z}(n).
\end{align}
For the exactly periodic case, the fidelity $\mathcal{Z}(m) = 1$ and the associated rate function $R(m)=0$, where $m$ is the period. In a dynamical sense, the fidelity measures the probability that the time-evolved state is found in the initial state. The rate function $R(n)$, on the other hand, plays a role similar to the free energy density in thermodynamics. Cusps in the rate function are interpreted as DQPTs. Our computations of the long-time-averaged rate function given by
\begin{align}
    \langle R \rangle = \lim_{n\to\infty} \frac{1}{n} \sum_{t = 0}^{n-1} R(t),
\end{align}
shows a pronounced peak at \(k_\theta = k_r\). However, for the case \(k_\theta = 0\), sharp changes in the rate function are observed only for half-odd integer values of \(j\).
\begin{figure}[!ht]
    \includegraphics[width=\linewidth]{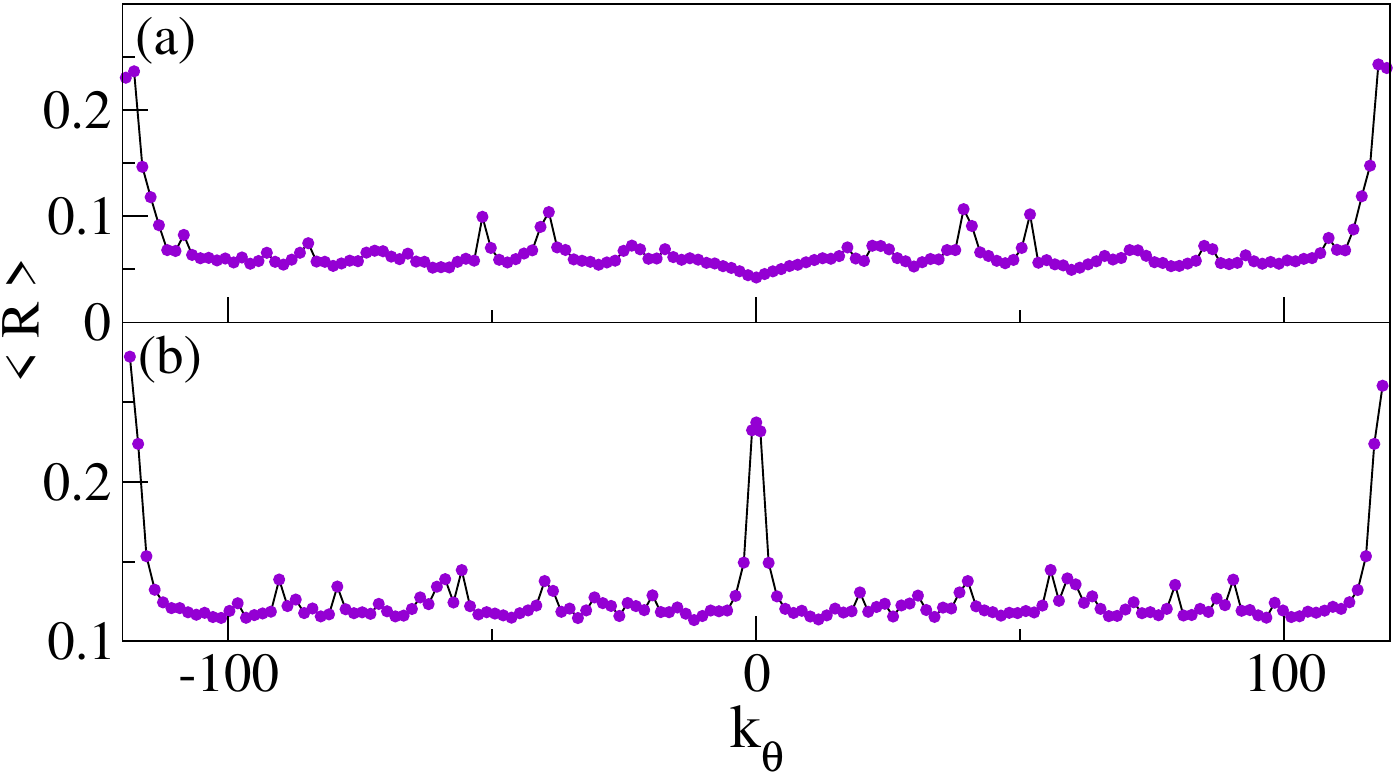}
    \caption{Average rate function \( \langle R \rangle \) obtained from the fidelity for the initial state \( |\theta_0 = 0, \phi_0 = 0\rangle \), evolved over \(n = 1000\) steps, with total spin (a) \(j = 76\) and (b) \(j = 75.5\). In both cases, \(k_r = j\pi/2\).}
    \label{figrate}
\end{figure}

\section{Behavior near $k_r = j\pi/2$}\label{sec:near_periodicity}
Absence of the semi-classical limit for the case $k_r = j\pi/2$, makes it fully quantum even in the limit $j\to\infty$. The periodicity of $\mathcal{U}$ for this case raises an important question: whether the DKT with $k_r = j\pi/2$ is quantum integrable or not~\citep{gritsev2017integrable}. We analyze the statistics of higher-order spacing ratios to support the integrability of the DKT near $k_r = j\pi/2$. 

\begin{figure}[!ht]
    \includegraphics[width=\linewidth]{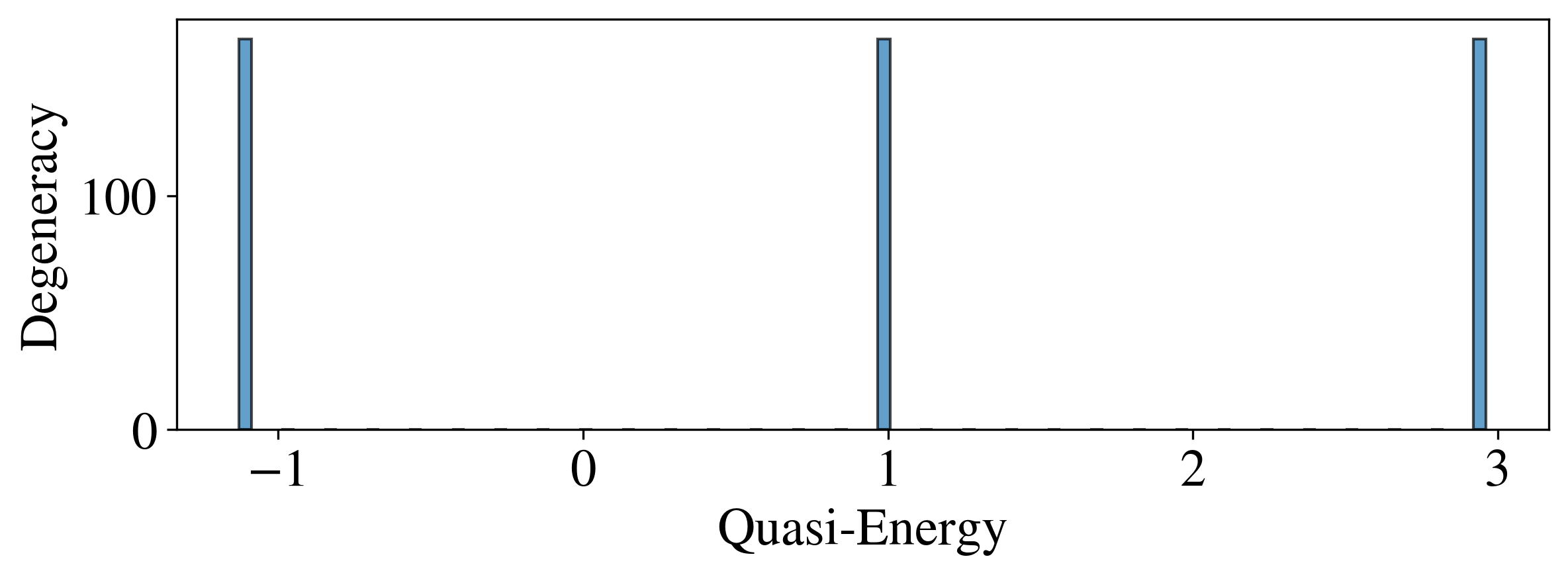}
    \caption{Degeneracy of the quasienergies of $\mathcal{U}$ for the DKT with \(p = \pi/2\), \( k_\theta = 0 \), \(k_r = j\pi/2\), and \(j = 500.5\).}\label{quasi-energy}
\end{figure}

\begin{figure}[!ht]
    \includegraphics[width=\linewidth]{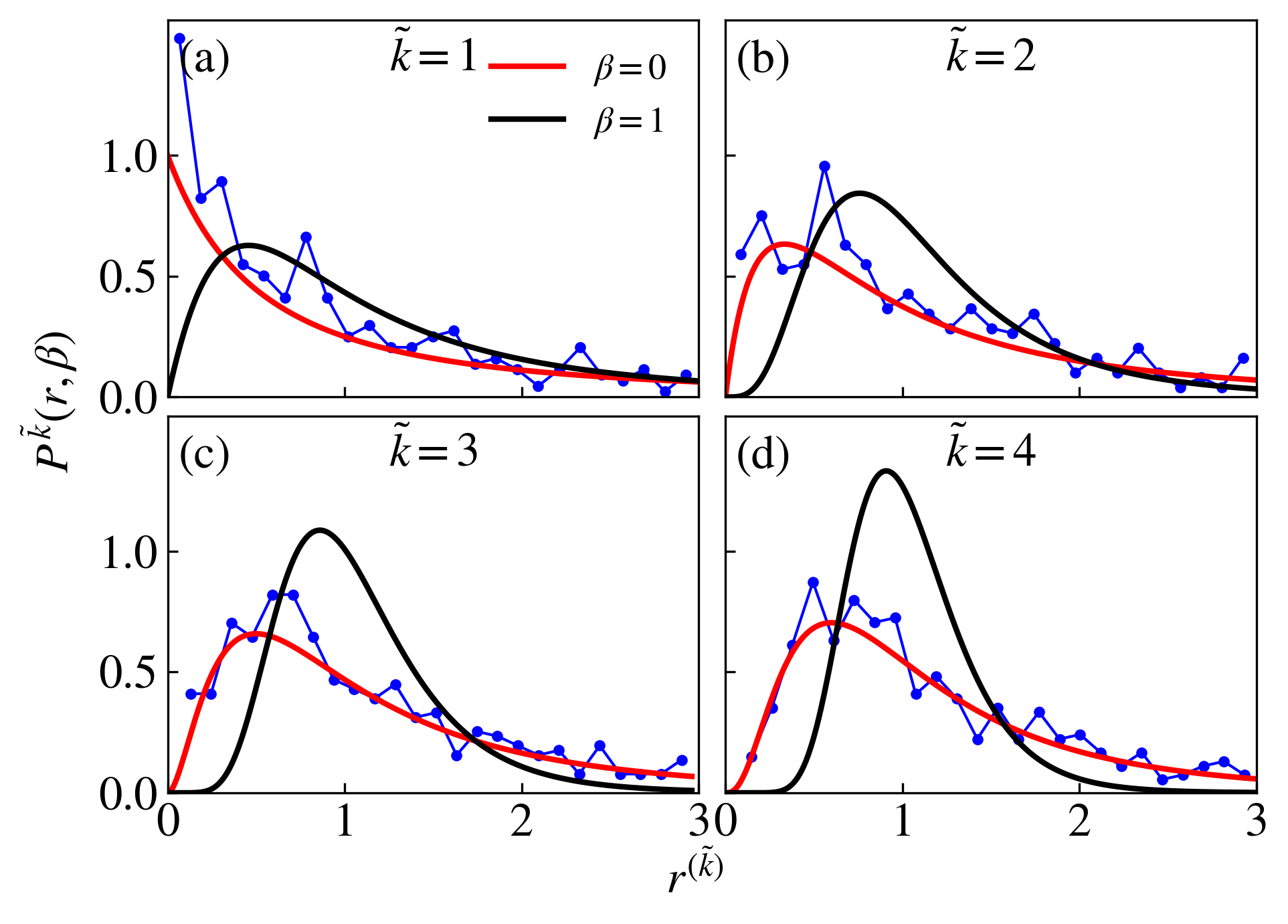}
    \caption{The probability distribution \(P^{\tilde{k}}(r, \beta)\) versus \(r^{(\tilde{k})}\) plotted for the DKT with \(p = \pi/2\), \( k_\theta = 0 \), \(k_r = 1.001 j\pi/2\), and \(j = 500.5\). Panels (a)--(d) correspond to order of $\tilde{k}$ from 1 to 4 respectively.}
    \label{RMT-poisson}
\end{figure}

\begin{figure}[!ht]
    \includegraphics[width=\linewidth]{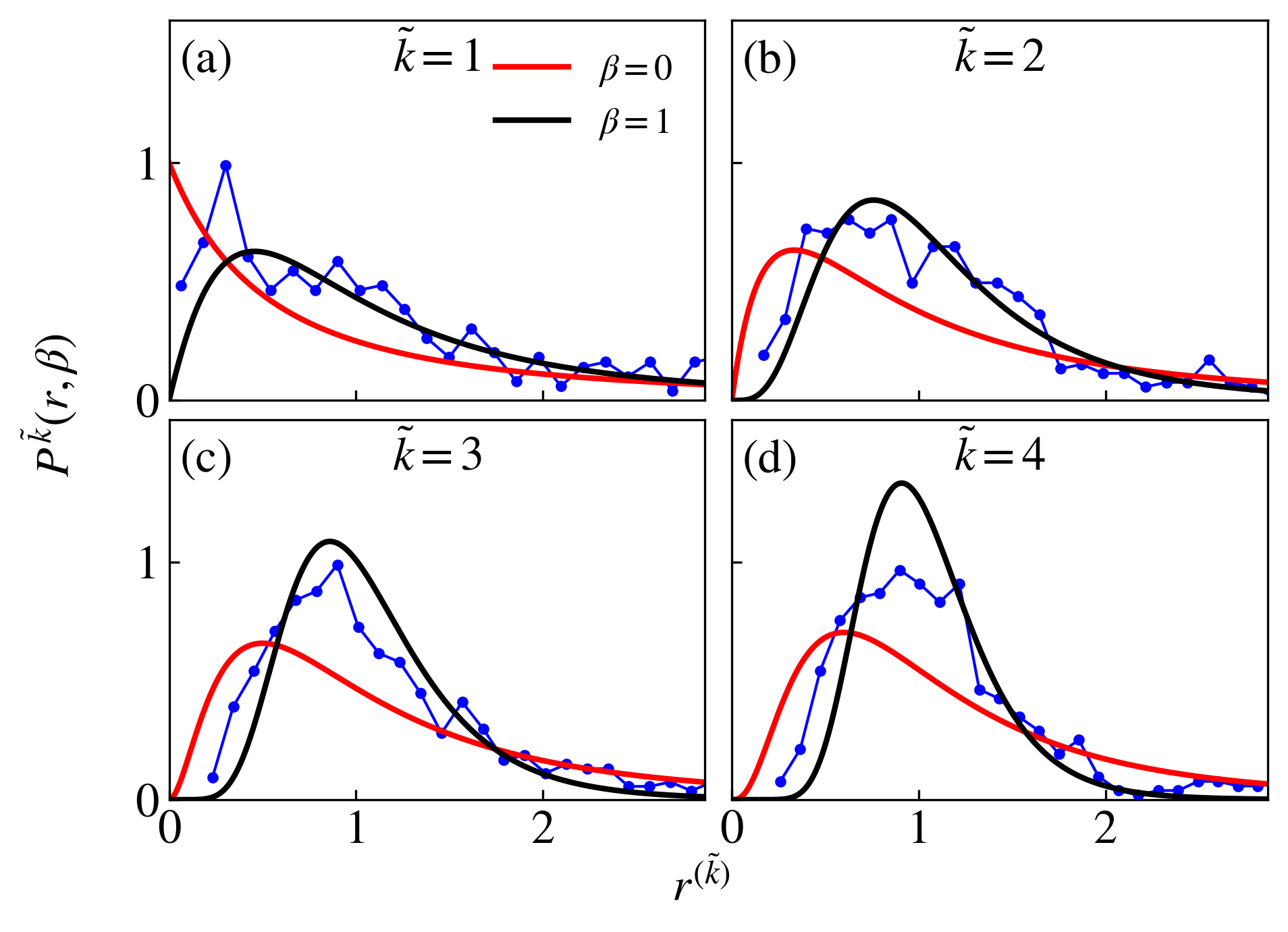}
    \caption{The probability distribution \(P^{\tilde{k}}(r, \beta)\) versus \(r^{(\tilde{k})}\) plotted for the DKT with \(p = \pi/2\), \( k_\theta = 0 \), \(k_r = 1.0015 j\pi/2\), and \(j = 500.5\). Panels (a)--(d) correspond to order of $\tilde{k}$ from 1 to 4 respectively.}
    \label{RMT-mixed}
\end{figure}

\begin{figure}[!ht]
    \includegraphics[width=\linewidth]{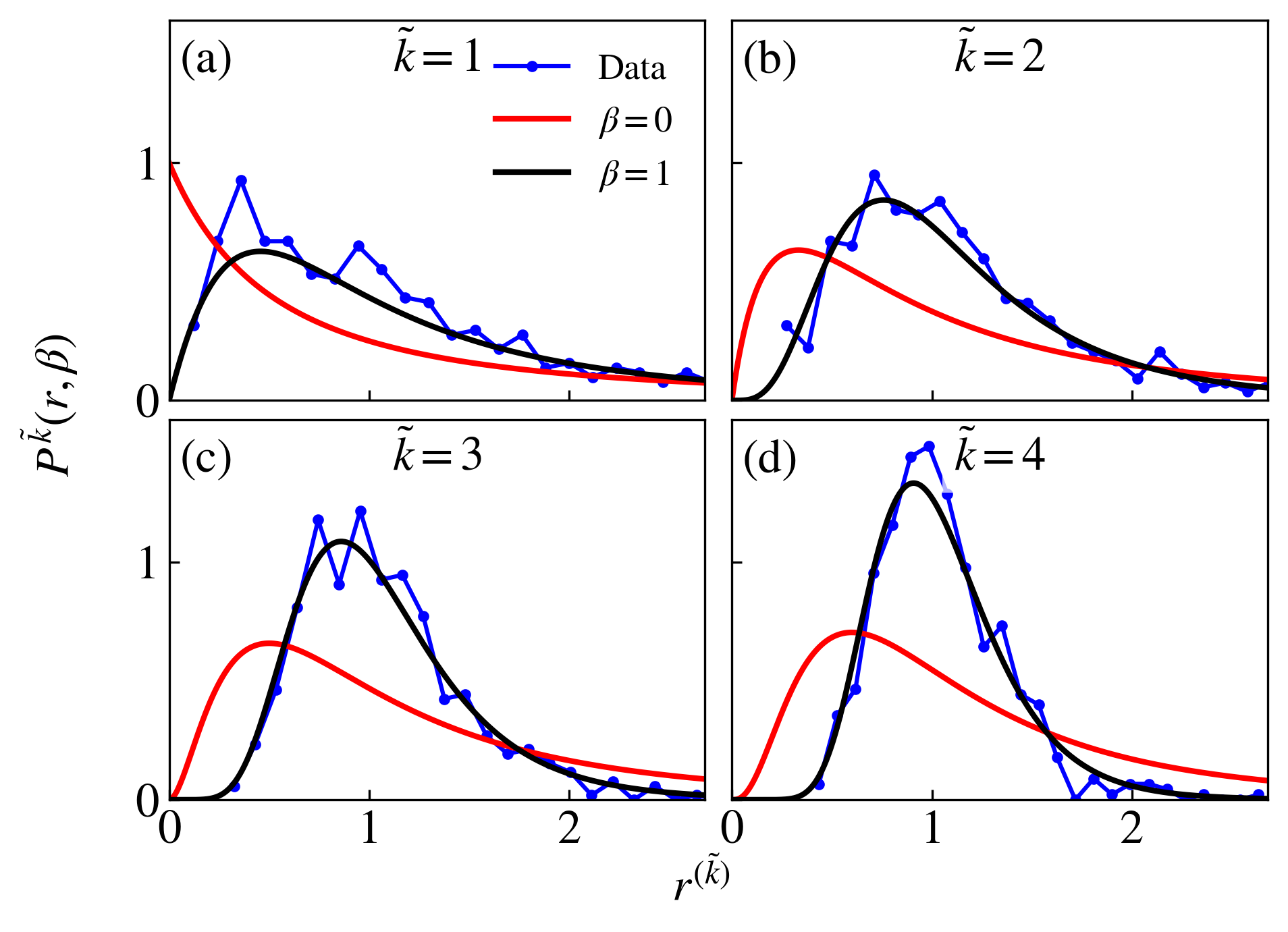}
    \caption{The probability distribution \(P^{\tilde{k}}(r, \beta)\) versus \(r^{(\tilde{k})}\) plotted for the DKT with \(p = \pi/2\), \( k_\theta = 0 \), \(k_r = 1.002 j\pi/2\), and \(j = 500.5\). Panels (a)--(d) correspond to order of $\tilde{k}$ from 1 to 4 respectively.}
    \label{RMT-nonintegrable}
\end{figure}

\begin{figure}[!ht]
    \includegraphics[width=\linewidth]{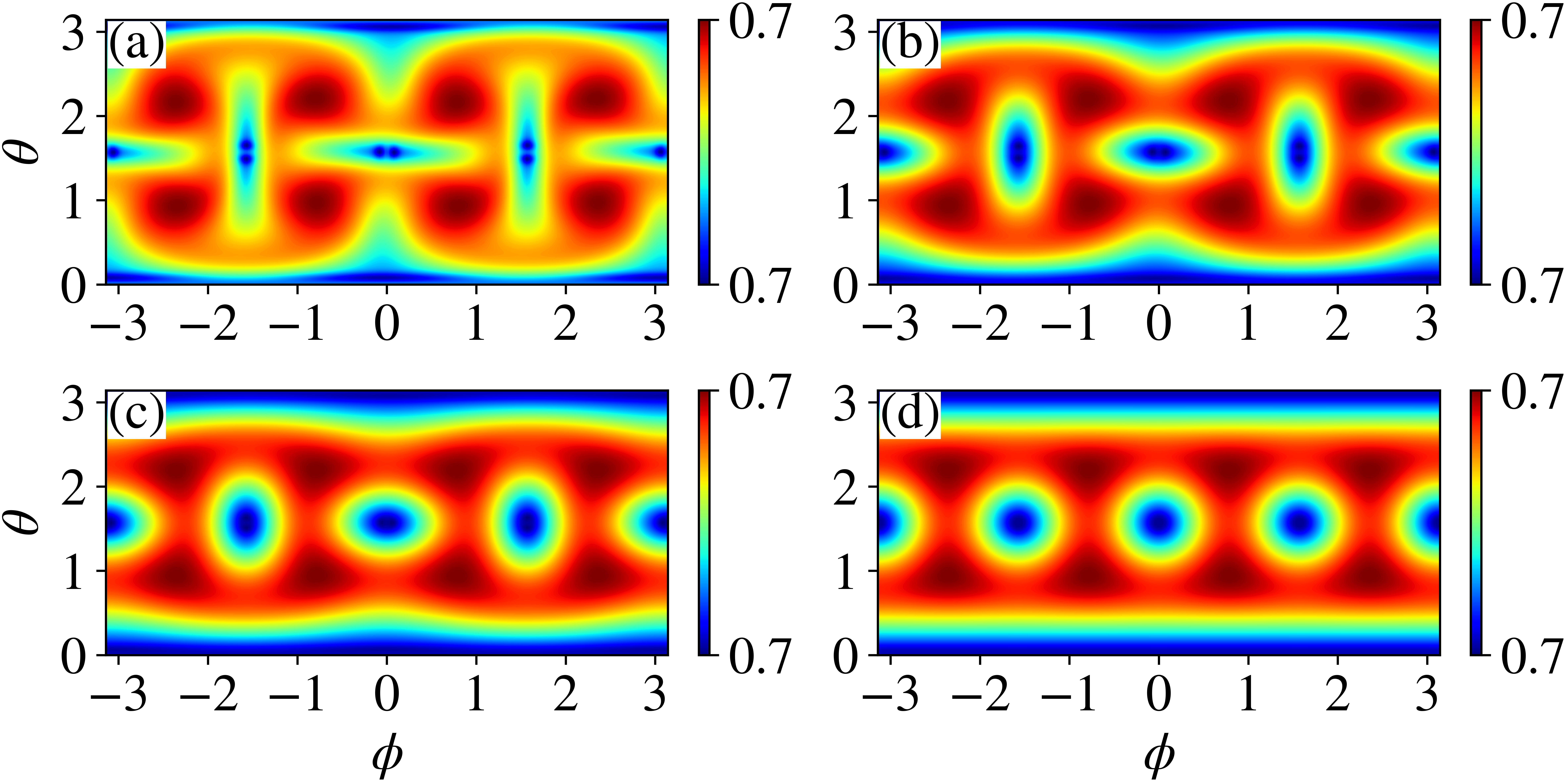}
    \caption{Numerically computed long-time-averaged von Neumann entropy of the single-particle reduced density matrix \(\rho_1(n)\) for total spin \(j = 75.5\). Here, 40,000 initial spin-coherent states are evolved over \(n = 1000\) steps. The special case \(k_\theta = k_r\) is considered for (a) \(k_r = 0.9996 j\pi/2\), (b) \(k_r = 0.9997 j\pi/2\), (c) \(k_r = 0.9998 j\pi/2\) and (d) \(k_r = j\pi/2\).}
    \label{near_periodicity}
\end{figure}

The spacing-ratio statistics (SRF) $P^{\tilde{k}}(r, \beta )$ is plotted as a function of the level-spacing ratio $r^{(\tilde{k})} = (E_{i+2k} - E_{i+k})/(E_{i+k} - E_i)$~\citep{mehta2004random,harshini2020symmetry,bhosale2028higher}, where $\lbrace E_i\rbrace$ represent quasi-energies corresponding to the Floquet operator $\mathcal{U}$. The distribution shows a highly degenerate spectrum at $k_r = j\pi/2$ (see Fig.~\ref{quasi-energy}). The addition of a small perturbation reveals a Poisson distribution as an indication of quantum integrability (see Fig.~\ref{RMT-poisson}). As $k_r$ is slightly increased, the SRF evolves into an intermediate statistics of Poisson and Gaussian Orthogonal Ensemble (GOE) (see Fig.~\ref{RMT-mixed}). With further increase in $k_r$, the SRF approaches the GOE behavior characteristic of non-integrable dynamics (see Fig.~\ref{RMT-nonintegrable}). This transition from integrable to non-integrable dynamics becomes sharper with increasing $j$. Nonetheless, for finite $j$, we can expect a mixed regime due to the presence of intermediate statistics for certain values of $k_r$ near $j\pi/2$. 
We now consider long-time-averaged von Neumann entropy to analyze the intermediate regime (see Fig.~\ref{near_periodicity}). Near $k_r = j\pi/2$, pairs of low-valued blue regions of long-time-averaged von Neumann entropy form near states \(|\theta_0 = 0, \phi_0 = 0\rangle\), \(|\theta_0 = \pi, \phi_0 = \pi\rangle\), \(|\theta_0 = \pi/2, \phi_0 = 0\rangle\), \(|\theta_0 = \pi/2, \phi_0 = \pi\rangle\) and \(|\theta_0 = \pi/2, \phi_0 = \pm \pi/2\rangle\) and merge as $k_r$ approaches $j\pi/2$ (see Figs.~\ref{near_periodicity}(a-d)). The corresponding SRF indicates the existence of a mixed regime—partially chaotic and partially regular in the quantum sense—of the order of $10^{-3}$ on both sides of $k_r = j\pi/2$ (see Fig.~\ref{RMT-mixed}).

This bifurcation phenomenon in the von Neumann entropy is noteworthy, as it is driven entirely by quantum effects. For large $k_r$, the DKT dynamics are classical and chaotic. But at $k_r = j\pi/2$, it is quantum and integrable. This shows how quantum effects can suppress classical chaos. 

\section{Results and Conclusion}\label{sec:rc}
In this work, we have investigated exact recurrences of the DKT using analytical and numerical techniques. Our key result is the identification of the exact periodicity of the Floquet operator for \(k_r = j\pi/2, j\pi/4\), and its independence from the symmetry-breaking parameter \(k_\theta\). At the exact quantum recurrence points, our computations of the Husimi distribution show that the quantum wave packets return to their initial position and become localized after a predicted time, regardless of the value of \(k_\theta\).

For integer \(j\), we find that the Floquet operator exhibits periodicity with period 8 when \(k_r = j\pi/2\), and period 48 when \(k_r = j\pi/4\). In contrast, for half-odd integer values of \(j\), the periodicity turns out to be 12 for \(k_r = j\pi/2\), but absence thereof for \(k_r = j\pi/4\). In all cases, the periodicity is shown to be independent of \(k_\theta\), despite its role in breaking temporal symmetry. With this, we have generalized the recent work on quantum recurrences \citep{amit2024} by explicit analytical derivations. 

To probe entanglement dynamics, we have studied the long-time-averaged von Neumann entropy of the single-qubit reduced density matrix. Breaking time-reversal symmetry by perturbing $k_\theta$ does not show any noteworthy change in entanglement for most initial states near $k_\theta = 0$. However, near $k_\theta = \pm k_r$, such perturbations create vortices. Motivated by these findings, we further examine the dynamical response of the initial state $|\theta_0 = 0, \phi_0 = 0\rangle$ using the fidelity rate function as a probe for DQPTs. Pronounced signatures are observed at \(k_\theta = \pm k_r\). However, for \(k_\theta = 0\), these features depend on whether $2j$ is even or odd. These results suggest a qualitative distinction between the two time-reversal symmetric cases $k_\theta = 0$ and $k_\theta = \pm k_r$. It is an open question whether the distinction between these two cases arises from local or global quantum dynamics. 

In the vicinity of the periodic point \(k_r = j\pi/2\), we observe signatures of integrable behavior in both entropy and level spacing ratio statistics. As \(k_r\) departs from \(j\pi/2\), the system transitions from integrable to non-integrable, and the states corresponding to the trivial fixed points exhibit bifurcation in the long-time-averaged von Neumann entropy. This bifurcation is unique as it does not have a classical analogue.

The exact recurrences identified in this work are useful from a quantum control perspective. Even though the transformed kick strengths \((k_r, k_\theta)\) are relevant for the dynamics, the physical kicks are \((k, k')\). This allows us to tune the system of any size to get a desirable output in terms of entanglement dynamics. 

The DKT can be realized by extending established QKT platforms. In NMR systems~\citep{krithika2023nmr,krithika2022quantum}, the DKT extension can be achieved by adding a second nonlinear twist  immediately after the first. In cold-atom setups~\citep{chaudhary_quantum_signatures} and Floquet synthetic lattice systems~\citep{meier2019exploring}, a second Stark-shift kick can be applied in a rotated basis. The same effect can be obtained using an additional quadratic-phase gate in the superconducting qubits~\citep{neil2016ergodic}.

In summary, our study reveals that exact quantum recurrences in the DKT are robust to the time-reversal symmetry-breaking parameter $k_\theta$. The DQPT is observed at \(k_\theta = \pm k_r\), but in the case of \(k_\theta = 0\), it depends on whether \(2j\) is even or odd, indicating a qualitative distinction between these two time-reversal symmetric cases. The transition from integrable to chaotic behavior at \(k_r = j\pi/2\) shows how quantum effects can suppress the underlying chaotic dynamics. These findings establish the DKT as a versatile platform for exploring quantum recurrences, effects of time-reversal symmetry, and controlling driven Floquet systems.

\section{Acknowledgments}
The authors are grateful to the Department of Science and Technology (DST) for their generous financial support, making this research possible through sanctioned Project No. SR/FST/PSI/2017/5(C) to the Department of Physics of VNIT, Nagpur.

\bibliographystyle{apsrev4-2}
\bibliography{ref}

\begin{thebibliography}{55}%
\makeatletter
\providecommand \@ifxundefined [1]{%
 \@ifx{#1\undefined}
}%
\providecommand \@ifnum [1]{%
 \ifnum #1\expandafter \@firstoftwo
 \else \expandafter \@secondoftwo
 \fi
}%
\providecommand \@ifx [1]{%
 \ifx #1\expandafter \@firstoftwo
 \else \expandafter \@secondoftwo
 \fi
}%
\providecommand \natexlab [1]{#1}%
\providecommand \enquote  [1]{``#1''}%
\providecommand \bibnamefont  [1]{#1}%
\providecommand \bibfnamefont [1]{#1}%
\providecommand \citenamefont [1]{#1}%
\providecommand \href@noop [0]{\@secondoftwo}%
\providecommand \href [0]{\begingroup \@sanitize@url \@href}%
\providecommand \@href[1]{\@@startlink{#1}\@@href}%
\providecommand \@@href[1]{\endgroup#1\@@endlink}%
\providecommand \@sanitize@url [0]{\catcode `\\12\catcode `\$12\catcode `\&12\catcode `\#12\catcode `\^12\catcode `\_12\catcode `\%12\relax}%
\providecommand \@@startlink[1]{}%
\providecommand \@@endlink[0]{}%
\providecommand \url  [0]{\begingroup\@sanitize@url \@url }%
\providecommand \@url [1]{\endgroup\@href {#1}{\urlprefix }}%
\providecommand \urlprefix  [0]{URL }%
\providecommand \Eprint [0]{\href }%
\providecommand \doibase [0]{https://doi.org/}%
\providecommand \selectlanguage [0]{\@gobble}%
\providecommand \bibinfo  [0]{\@secondoftwo}%
\providecommand \bibfield  [0]{\@secondoftwo}%
\providecommand \translation [1]{[#1]}%
\providecommand \BibitemOpen [0]{}%
\providecommand \bibitemStop [0]{}%
\providecommand \bibitemNoStop [0]{.\EOS\space}%
\providecommand \EOS [0]{\spacefactor3000\relax}%
\providecommand \BibitemShut  [1]{\csname bibitem#1\endcsname}%
\let\auto@bib@innerbib\@empty
\bibitem [{\citenamefont {Barreira}(2006)}]{barreira2006poincare}%
  \BibitemOpen
  \bibfield  {author} {\bibinfo {author} {\bibfnamefont {L.}~\bibnamefont {Barreira}},\ }in\ \href@noop {} {\emph {\bibinfo {booktitle} {XIVth International Congress on Mathematical Physics}}}\ (\bibinfo {organization} {World Scientific},\ \bibinfo {year} {2006})\ pp.\ \bibinfo {pages} {415--422}\BibitemShut {NoStop}%
\bibitem [{\citenamefont {Anishchenko}\ and\ \citenamefont {Astakhov}(2013)}]{anishchenko2013poincare}%
  \BibitemOpen
  \bibfield  {author} {\bibinfo {author} {\bibfnamefont {V.~S.}\ \bibnamefont {Anishchenko}}\ and\ \bibinfo {author} {\bibfnamefont {S.~V.}\ \bibnamefont {Astakhov}},\ }\href@noop {} {\bibfield  {journal} {\bibinfo  {journal} {Physics-Uspekhi}\ }\textbf {\bibinfo {volume} {56}},\ \bibinfo {pages} {955} (\bibinfo {year} {2013})}\BibitemShut {NoStop}%
\bibitem [{\citenamefont {Saussol}(2009)}]{saussol2009introduction}%
  \BibitemOpen
  \bibfield  {author} {\bibinfo {author} {\bibfnamefont {B.}~\bibnamefont {Saussol}},\ }\href@noop {} {\bibfield  {journal} {\bibinfo  {journal} {Reviews in Mathematical Physics}\ }\textbf {\bibinfo {volume} {21}},\ \bibinfo {pages} {949} (\bibinfo {year} {2009})}\BibitemShut {NoStop}%
\bibitem [{\citenamefont {Bocchieri}\ and\ \citenamefont {Loinger}(1957)}]{bocchieri1957quantum}%
  \BibitemOpen
  \bibfield  {author} {\bibinfo {author} {\bibfnamefont {P.}~\bibnamefont {Bocchieri}}\ and\ \bibinfo {author} {\bibfnamefont {A.}~\bibnamefont {Loinger}},\ }\href@noop {} {\bibfield  {journal} {\bibinfo  {journal} {Phys. Rev.}\ }\textbf {\bibinfo {volume} {107}},\ \bibinfo {pages} {337} (\bibinfo {year} {1957})}\BibitemShut {NoStop}%
\bibitem [{\citenamefont {Peres}(1982)}]{peres1982recurrence}%
  \BibitemOpen
  \bibfield  {author} {\bibinfo {author} {\bibfnamefont {A.}~\bibnamefont {Peres}},\ }\href@noop {} {\bibfield  {journal} {\bibinfo  {journal} {Phys. Rev. Lett.}\ }\textbf {\bibinfo {volume} {49}},\ \bibinfo {pages} {1118} (\bibinfo {year} {1982})}\BibitemShut {NoStop}%
\bibitem [{\citenamefont {Schulman}(1978)}]{schulman1978note}%
  \BibitemOpen
  \bibfield  {author} {\bibinfo {author} {\bibfnamefont {L.~S.}\ \bibnamefont {Schulman}},\ }\href@noop {} {\bibfield  {journal} {\bibinfo  {journal} {Phys. Rev. A}\ }\textbf {\bibinfo {volume} {18}},\ \bibinfo {pages} {2379} (\bibinfo {year} {1978})}\BibitemShut {NoStop}%
\bibitem [{\citenamefont {Fishman}\ \emph {et~al.}(1982)\citenamefont {Fishman}, \citenamefont {Grempel},\ and\ \citenamefont {Prange}}]{fishman1982chaos}%
  \BibitemOpen
  \bibfield  {author} {\bibinfo {author} {\bibfnamefont {S.}~\bibnamefont {Fishman}}, \bibinfo {author} {\bibfnamefont {D.}~\bibnamefont {Grempel}},\ and\ \bibinfo {author} {\bibfnamefont {R.~E.}\ \bibnamefont {Prange}},\ }\href@noop {} {\bibfield  {journal} {\bibinfo  {journal} {Phys. Rev. Lett.}\ }\textbf {\bibinfo {volume} {49}},\ \bibinfo {pages} {509} (\bibinfo {year} {1982})}\BibitemShut {NoStop}%
\bibitem [{\citenamefont {Wigner}(1951)}]{wigner1951statistical}%
  \BibitemOpen
  \bibfield  {author} {\bibinfo {author} {\bibfnamefont {E.~P.}\ \bibnamefont {Wigner}},\ }in\ \href@noop {} {\emph {\bibinfo {booktitle} {Mathematical Proceedings of the Cambridge Philosophical Society}}},\ Vol.~\bibinfo {volume} {47}\ (\bibinfo {organization} {Cambridge University Press},\ \bibinfo {year} {1951})\ pp.\ \bibinfo {pages} {790--798}\BibitemShut {NoStop}%
\bibitem [{\citenamefont {Dyson}(1962)}]{dyson1962statistical}%
  \BibitemOpen
  \bibfield  {author} {\bibinfo {author} {\bibfnamefont {F.~J.}\ \bibnamefont {Dyson}},\ }\href@noop {} {\bibfield  {journal} {\bibinfo  {journal} {Journal of Mathematical Physics}\ }\textbf {\bibinfo {volume} {3}},\ \bibinfo {pages} {140} (\bibinfo {year} {1962})}\BibitemShut {NoStop}%
\bibitem [{\citenamefont {Berry}\ and\ \citenamefont {Tabor}(1977)}]{berry1977level}%
  \BibitemOpen
  \bibfield  {author} {\bibinfo {author} {\bibfnamefont {M.~V.}\ \bibnamefont {Berry}}\ and\ \bibinfo {author} {\bibfnamefont {M.}~\bibnamefont {Tabor}},\ }\href@noop {} {\bibfield  {journal} {\bibinfo  {journal} {Proceedings of the Royal Society of London. A. Mathematical and Physical Sciences}\ }\textbf {\bibinfo {volume} {356}},\ \bibinfo {pages} {375} (\bibinfo {year} {1977})}\BibitemShut {NoStop}%
\bibitem [{\citenamefont {Bohigas}\ \emph {et~al.}(1984)\citenamefont {Bohigas}, \citenamefont {Giannoni},\ and\ \citenamefont {Schmit}}]{bohigas1984characterization}%
  \BibitemOpen
  \bibfield  {author} {\bibinfo {author} {\bibfnamefont {O.}~\bibnamefont {Bohigas}}, \bibinfo {author} {\bibfnamefont {M.-J.}\ \bibnamefont {Giannoni}},\ and\ \bibinfo {author} {\bibfnamefont {C.}~\bibnamefont {Schmit}},\ }\href@noop {} {\bibfield  {journal} {\bibinfo  {journal} {Phys. Rev. Lett.}\ }\textbf {\bibinfo {volume} {52}},\ \bibinfo {pages} {1} (\bibinfo {year} {1984})}\BibitemShut {NoStop}%
\bibitem [{\citenamefont {Korenblit}\ \emph {et~al.}(2012)\citenamefont {Korenblit}, \citenamefont {Kafri}, \citenamefont {Campbell}, \citenamefont {Islam}, \citenamefont {Edwards}, \citenamefont {Gong}, \citenamefont {Lin}, \citenamefont {Duan}, \citenamefont {Kim}, \citenamefont {Kim} \emph {et~al.}}]{korenblit2012arbitrary}%
  \BibitemOpen
  \bibfield  {author} {\bibinfo {author} {\bibfnamefont {S.}~\bibnamefont {Korenblit}}, \bibinfo {author} {\bibfnamefont {D.}~\bibnamefont {Kafri}}, \bibinfo {author} {\bibfnamefont {W.~C.}\ \bibnamefont {Campbell}}, \bibinfo {author} {\bibfnamefont {R.}~\bibnamefont {Islam}}, \bibinfo {author} {\bibfnamefont {E.~E.}\ \bibnamefont {Edwards}}, \bibinfo {author} {\bibfnamefont {Z.-X.}\ \bibnamefont {Gong}}, \bibinfo {author} {\bibfnamefont {G.-D.}\ \bibnamefont {Lin}}, \bibinfo {author} {\bibfnamefont {L.-M.}\ \bibnamefont {Duan}}, \bibinfo {author} {\bibfnamefont {J.}~\bibnamefont {Kim}}, \bibinfo {author} {\bibfnamefont {K.}~\bibnamefont {Kim}}, \emph {et~al.},\ }\href@noop {} {\bibfield  {journal} {\bibinfo  {journal} {New Journal of Physics}\ }\textbf {\bibinfo {volume} {14}},\ \bibinfo {pages} {095024} (\bibinfo {year} {2012})}\BibitemShut {NoStop}%
\bibitem [{\citenamefont {Peres}(1984{\natexlab{a}})}]{peres1984stability}%
  \BibitemOpen
  \bibfield  {author} {\bibinfo {author} {\bibfnamefont {A.}~\bibnamefont {Peres}},\ }\href@noop {} {\bibfield  {journal} {\bibinfo  {journal} {Phys. Rev. A}\ }\textbf {\bibinfo {volume} {30}},\ \bibinfo {pages} {1610} (\bibinfo {year} {1984}{\natexlab{a}})}\BibitemShut {NoStop}%
\bibitem [{\citenamefont {Peres}(1984{\natexlab{b}})}]{peres1984ergodicity}%
  \BibitemOpen
  \bibfield  {author} {\bibinfo {author} {\bibfnamefont {A.}~\bibnamefont {Peres}},\ }\href@noop {} {\bibfield  {journal} {\bibinfo  {journal} {Phys. Rev. A}\ }\textbf {\bibinfo {volume} {30}},\ \bibinfo {pages} {504} (\bibinfo {year} {1984}{\natexlab{b}})}\BibitemShut {NoStop}%
\bibitem [{\citenamefont {Feingold}\ \emph {et~al.}(1984)\citenamefont {Feingold}, \citenamefont {Moiseyev},\ and\ \citenamefont {Peres}}]{feingold1984ergodicity}%
  \BibitemOpen
  \bibfield  {author} {\bibinfo {author} {\bibfnamefont {M.}~\bibnamefont {Feingold}}, \bibinfo {author} {\bibfnamefont {N.}~\bibnamefont {Moiseyev}},\ and\ \bibinfo {author} {\bibfnamefont {A.}~\bibnamefont {Peres}},\ }\href@noop {} {\bibfield  {journal} {\bibinfo  {journal} {Phys. Rev. A}\ }\textbf {\bibinfo {volume} {30}},\ \bibinfo {pages} {509} (\bibinfo {year} {1984})}\BibitemShut {NoStop}%
\bibitem [{\citenamefont {Haake}\ \emph {et~al.}(1987)\citenamefont {Haake}, \citenamefont {Ku{\'s}},\ and\ \citenamefont {Scharf}}]{haake1987classical}%
  \BibitemOpen
  \bibfield  {author} {\bibinfo {author} {\bibfnamefont {F.}~\bibnamefont {Haake}}, \bibinfo {author} {\bibfnamefont {M.}~\bibnamefont {Ku{\'s}}},\ and\ \bibinfo {author} {\bibfnamefont {R.}~\bibnamefont {Scharf}},\ }\href@noop {} {\bibfield  {journal} {\bibinfo  {journal} {Zeitschrift f{\"u}r Physik B Condensed Matter}\ }\textbf {\bibinfo {volume} {65}},\ \bibinfo {pages} {381} (\bibinfo {year} {1987})}\BibitemShut {NoStop}%
\bibitem [{\citenamefont {Chaudhury}\ \emph {et~al.}(2009)\citenamefont {Chaudhury}, \citenamefont {Smith}, \citenamefont {Anderson}, \citenamefont {Ghose},\ and\ \citenamefont {Jessen}}]{chaudhary_quantum_signatures}%
  \BibitemOpen
  \bibfield  {author} {\bibinfo {author} {\bibfnamefont {S.}~\bibnamefont {Chaudhury}}, \bibinfo {author} {\bibfnamefont {A.}~\bibnamefont {Smith}}, \bibinfo {author} {\bibfnamefont {B.}~\bibnamefont {Anderson}}, \bibinfo {author} {\bibfnamefont {S.}~\bibnamefont {Ghose}},\ and\ \bibinfo {author} {\bibfnamefont {P.~S.}\ \bibnamefont {Jessen}},\ }\href@noop {} {\bibfield  {journal} {\bibinfo  {journal} {Nature}\ }\textbf {\bibinfo {volume} {461}},\ \bibinfo {pages} {768} (\bibinfo {year} {2009})}\BibitemShut {NoStop}%
\bibitem [{\citenamefont {Krithika}\ \emph {et~al.}(2023)\citenamefont {Krithika}, \citenamefont {Santhanam},\ and\ \citenamefont {Mahesh}}]{krithika2023nmr}%
  \BibitemOpen
  \bibfield  {author} {\bibinfo {author} {\bibfnamefont {V.~R.}\ \bibnamefont {Krithika}}, \bibinfo {author} {\bibfnamefont {M.~S.}\ \bibnamefont {Santhanam}},\ and\ \bibinfo {author} {\bibfnamefont {T.~S.}\ \bibnamefont {Mahesh}},\ }\href@noop {} {\bibfield  {journal} {\bibinfo  {journal} {Phys. Rev. A}\ }\textbf {\bibinfo {volume} {108}},\ \bibinfo {pages} {032207} (\bibinfo {year} {2023})}\BibitemShut {NoStop}%
\bibitem [{\citenamefont {Krithika}\ \emph {et~al.}(2019)\citenamefont {Krithika}, \citenamefont {Anjusha}, \citenamefont {Bhosale},\ and\ \citenamefont {Mahesh}}]{krithika2022quantum}%
  \BibitemOpen
  \bibfield  {author} {\bibinfo {author} {\bibfnamefont {V.~R.}\ \bibnamefont {Krithika}}, \bibinfo {author} {\bibfnamefont {V.~S.}\ \bibnamefont {Anjusha}}, \bibinfo {author} {\bibfnamefont {U.~T.}\ \bibnamefont {Bhosale}},\ and\ \bibinfo {author} {\bibfnamefont {T.~S.}\ \bibnamefont {Mahesh}},\ }\href@noop {} {\bibfield  {journal} {\bibinfo  {journal} {Phys. Rev. E}\ }\textbf {\bibinfo {volume} {99}},\ \bibinfo {pages} {032219} (\bibinfo {year} {2019})}\BibitemShut {NoStop}%
\bibitem [{\citenamefont {Neill}\ \emph {et~al.}(2016)\citenamefont {Neill}, \citenamefont {Roushan}, \citenamefont {Fang}, \citenamefont {Chen}, \citenamefont {Kolodrubetz}, \citenamefont {Chen}, \citenamefont {Megrant}, \citenamefont {Barends}, \citenamefont {Campbell}, \citenamefont {Chiaro}, \citenamefont {Dunsworth}, \citenamefont {Jeffrey}, \citenamefont {Kelly}, \citenamefont {Mutus}, \citenamefont {O'Malley}, \citenamefont {Quintana}, \citenamefont {Sank}, \citenamefont {Vainsencher}, \citenamefont {Wenner}, \citenamefont {White}, \citenamefont {Polkovnikov},\ and\ \citenamefont {Martinis}}]{neil2016ergodic}%
  \BibitemOpen
  \bibfield  {author} {\bibinfo {author} {\bibfnamefont {C.}~\bibnamefont {Neill}}, \bibinfo {author} {\bibfnamefont {P.}~\bibnamefont {Roushan}}, \bibinfo {author} {\bibfnamefont {M.}~\bibnamefont {Fang}}, \bibinfo {author} {\bibfnamefont {Y.}~\bibnamefont {Chen}}, \bibinfo {author} {\bibfnamefont {M.}~\bibnamefont {Kolodrubetz}}, \bibinfo {author} {\bibfnamefont {Z.}~\bibnamefont {Chen}}, \bibinfo {author} {\bibfnamefont {A.}~\bibnamefont {Megrant}}, \bibinfo {author} {\bibfnamefont {R.}~\bibnamefont {Barends}}, \bibinfo {author} {\bibfnamefont {B.}~\bibnamefont {Campbell}}, \bibinfo {author} {\bibfnamefont {B.}~\bibnamefont {Chiaro}}, \bibinfo {author} {\bibfnamefont {A.}~\bibnamefont {Dunsworth}}, \bibinfo {author} {\bibfnamefont {E.}~\bibnamefont {Jeffrey}}, \bibinfo {author} {\bibfnamefont {J.}~\bibnamefont {Kelly}}, \bibinfo {author} {\bibfnamefont {J.}~\bibnamefont {Mutus}}, \bibinfo {author} {\bibfnamefont {P.~J.~J.}\ \bibnamefont {O'Malley}}, \bibinfo {author} {\bibfnamefont {C.}~\bibnamefont {Quintana}}, \bibinfo {author} {\bibfnamefont {D.}~\bibnamefont {Sank}}, \bibinfo {author} {\bibfnamefont {A.}~\bibnamefont {Vainsencher}}, \bibinfo {author} {\bibfnamefont {J.}~\bibnamefont {Wenner}}, \bibinfo {author} {\bibfnamefont {T.~C.}\ \bibnamefont {White}}, \bibinfo {author} {\bibfnamefont {A.}~\bibnamefont {Polkovnikov}},\ and\ \bibinfo {author} {\bibfnamefont {J.~M.}\ \bibnamefont {Martinis}},\ }\href@noop {} {\bibfield  {journal} {\bibinfo  {journal} {Nature Physics}\ }\textbf {\bibinfo {volume} {12}},\ \bibinfo {pages} {1037} (\bibinfo {year} {2016})}\BibitemShut {NoStop}%
\bibitem [{\citenamefont {Monroe}\ \emph {et~al.}(2021)\citenamefont {Monroe}, \citenamefont {Campbell}, \citenamefont {Duan}, \citenamefont {Gong}, \citenamefont {Gorshkov}, \citenamefont {Hess}, \citenamefont {Islam}, \citenamefont {Kim}, \citenamefont {Linke}, \citenamefont {Pagano} \emph {et~al.}}]{monroe2021programmable}%
  \BibitemOpen
  \bibfield  {author} {\bibinfo {author} {\bibfnamefont {C.}~\bibnamefont {Monroe}}, \bibinfo {author} {\bibfnamefont {W.~C.}\ \bibnamefont {Campbell}}, \bibinfo {author} {\bibfnamefont {L.-M.}\ \bibnamefont {Duan}}, \bibinfo {author} {\bibfnamefont {Z.-X.}\ \bibnamefont {Gong}}, \bibinfo {author} {\bibfnamefont {A.~V.}\ \bibnamefont {Gorshkov}}, \bibinfo {author} {\bibfnamefont {P.~W.}\ \bibnamefont {Hess}}, \bibinfo {author} {\bibfnamefont {R.}~\bibnamefont {Islam}}, \bibinfo {author} {\bibfnamefont {K.}~\bibnamefont {Kim}}, \bibinfo {author} {\bibfnamefont {N.~M.}\ \bibnamefont {Linke}}, \bibinfo {author} {\bibfnamefont {G.}~\bibnamefont {Pagano}}, \emph {et~al.},\ }\href@noop {} {\bibfield  {journal} {\bibinfo  {journal} {Reviews of Modern Physics}\ }\textbf {\bibinfo {volume} {93}},\ \bibinfo {pages} {025001} (\bibinfo {year} {2021})}\BibitemShut {NoStop}%
\bibitem [{\citenamefont {Pezze}\ \emph {et~al.}(2018)\citenamefont {Pezze}, \citenamefont {Smerzi}, \citenamefont {Oberthaler}, \citenamefont {Schmied},\ and\ \citenamefont {Treutlein}}]{pezze2018quantum}%
  \BibitemOpen
  \bibfield  {author} {\bibinfo {author} {\bibfnamefont {L.}~\bibnamefont {Pezze}}, \bibinfo {author} {\bibfnamefont {A.}~\bibnamefont {Smerzi}}, \bibinfo {author} {\bibfnamefont {M.~K.}\ \bibnamefont {Oberthaler}}, \bibinfo {author} {\bibfnamefont {R.}~\bibnamefont {Schmied}},\ and\ \bibinfo {author} {\bibfnamefont {P.}~\bibnamefont {Treutlein}},\ }\href@noop {} {\bibfield  {journal} {\bibinfo  {journal} {Reviews of Modern Physics}\ }\textbf {\bibinfo {volume} {90}},\ \bibinfo {pages} {035005} (\bibinfo {year} {2018})}\BibitemShut {NoStop}%
\bibitem [{\citenamefont {Zhang}\ \emph {et~al.}(2007)\citenamefont {Zhang}, \citenamefont {Liu}, \citenamefont {Chen},\ and\ \citenamefont {Li}}]{zhang2008four}%
  \BibitemOpen
  \bibfield  {author} {\bibinfo {author} {\bibfnamefont {T.}~\bibnamefont {Zhang}}, \bibinfo {author} {\bibfnamefont {W.-T.}\ \bibnamefont {Liu}}, \bibinfo {author} {\bibfnamefont {P.-X.}\ \bibnamefont {Chen}},\ and\ \bibinfo {author} {\bibfnamefont {C.-Z.}\ \bibnamefont {Li}},\ }\href@noop {} {\bibfield  {journal} {\bibinfo  {journal} {Phys. Rev. A}\ }\textbf {\bibinfo {volume} {75}},\ \bibinfo {pages} {062102} (\bibinfo {year} {2007})}\BibitemShut {NoStop}%
\bibitem [{\citenamefont {Campa}\ \emph {et~al.}(2009)\citenamefont {Campa}, \citenamefont {Dauxois},\ and\ \citenamefont {Ruffo}}]{campa2009statistical}%
  \BibitemOpen
  \bibfield  {author} {\bibinfo {author} {\bibfnamefont {A.}~\bibnamefont {Campa}}, \bibinfo {author} {\bibfnamefont {T.}~\bibnamefont {Dauxois}},\ and\ \bibinfo {author} {\bibfnamefont {S.}~\bibnamefont {Ruffo}},\ }\href@noop {} {\bibfield  {journal} {\bibinfo  {journal} {Physics Reports}\ }\textbf {\bibinfo {volume} {480}},\ \bibinfo {pages} {57} (\bibinfo {year} {2009})}\BibitemShut {NoStop}%
\bibitem [{\citenamefont {Kastner}(2011)}]{kastner2011diverging}%
  \BibitemOpen
  \bibfield  {author} {\bibinfo {author} {\bibfnamefont {M.}~\bibnamefont {Kastner}},\ }\href@noop {} {\bibfield  {journal} {\bibinfo  {journal} {Phys. Rev. Lett.}\ }\textbf {\bibinfo {volume} {106}},\ \bibinfo {pages} {130601} (\bibinfo {year} {2011})}\BibitemShut {NoStop}%
\bibitem [{\citenamefont {Fey}\ and\ \citenamefont {Schmidt}(2016)}]{fey2016ising}%
  \BibitemOpen
  \bibfield  {author} {\bibinfo {author} {\bibfnamefont {S.}~\bibnamefont {Fey}}\ and\ \bibinfo {author} {\bibfnamefont {K.~P.}\ \bibnamefont {Schmidt}},\ }\href@noop {} {\bibfield  {journal} {\bibinfo  {journal} {Phys. Rev. B}\ }\textbf {\bibinfo {volume} {94}},\ \bibinfo {pages} {075156} (\bibinfo {year} {2016})}\BibitemShut {NoStop}%
\bibitem [{\citenamefont {Britton}\ \emph {et~al.}(2012)\citenamefont {Britton}, \citenamefont {Sawyer}, \citenamefont {Keith}, \citenamefont {Wang}, \citenamefont {Freericks}, \citenamefont {Uys}, \citenamefont {Biercuk},\ and\ \citenamefont {Bollinger}}]{britton2012engineered}%
  \BibitemOpen
  \bibfield  {author} {\bibinfo {author} {\bibfnamefont {J.~W.}\ \bibnamefont {Britton}}, \bibinfo {author} {\bibfnamefont {B.~C.}\ \bibnamefont {Sawyer}}, \bibinfo {author} {\bibfnamefont {A.~C.}\ \bibnamefont {Keith}}, \bibinfo {author} {\bibfnamefont {C.-C.~J.}\ \bibnamefont {Wang}}, \bibinfo {author} {\bibfnamefont {J.~K.}\ \bibnamefont {Freericks}}, \bibinfo {author} {\bibfnamefont {H.}~\bibnamefont {Uys}}, \bibinfo {author} {\bibfnamefont {M.~J.}\ \bibnamefont {Biercuk}},\ and\ \bibinfo {author} {\bibfnamefont {J.~J.}\ \bibnamefont {Bollinger}},\ }\href@noop {} {\bibfield  {journal} {\bibinfo  {journal} {Nature}\ }\textbf {\bibinfo {volume} {484}},\ \bibinfo {pages} {489} (\bibinfo {year} {2012})}\BibitemShut {NoStop}%
\bibitem [{\citenamefont {Peebles}(1987)}]{peebles1980large}%
  \BibitemOpen
  \bibfield  {author} {\bibinfo {author} {\bibfnamefont {P.}~\bibnamefont {Peebles}},\ }\href@noop {} {\bibfield  {journal} {\bibinfo  {journal} {Astrophysical Journal, Part 1 (ISSN 0004-637X), vol. 317, June 15, 1987, p. 576-587. NSF-supported research.}\ }\textbf {\bibinfo {volume} {317}},\ \bibinfo {pages} {576} (\bibinfo {year} {1987})}\BibitemShut {NoStop}%
\bibitem [{\citenamefont {Bouchet}\ \emph {et~al.}(2010)\citenamefont {Bouchet}, \citenamefont {Gupta},\ and\ \citenamefont {Mukamel}}]{bouchet2010thermodynamics}%
  \BibitemOpen
  \bibfield  {author} {\bibinfo {author} {\bibfnamefont {F.}~\bibnamefont {Bouchet}}, \bibinfo {author} {\bibfnamefont {S.}~\bibnamefont {Gupta}},\ and\ \bibinfo {author} {\bibfnamefont {D.}~\bibnamefont {Mukamel}},\ }\href@noop {} {\bibfield  {journal} {\bibinfo  {journal} {Physica A: Statistical Mechanics and its Applications}\ }\textbf {\bibinfo {volume} {389}},\ \bibinfo {pages} {4389} (\bibinfo {year} {2010})}\BibitemShut {NoStop}%
\bibitem [{\citenamefont {Bohr}\ and\ \citenamefont {Mottelson}(1998)}]{bohr1998nuclear}%
  \BibitemOpen
  \bibfield  {author} {\bibinfo {author} {\bibfnamefont {A.}~\bibnamefont {Bohr}}\ and\ \bibinfo {author} {\bibfnamefont {B.~R.}\ \bibnamefont {Mottelson}},\ }\href@noop {} {\emph {\bibinfo {title} {Nuclear structure}}},\ Vol.~\bibinfo {volume} {1}\ (\bibinfo  {publisher} {World scientific},\ \bibinfo {year} {1998})\BibitemShut {NoStop}%
\bibitem [{\citenamefont {Brink}(1993)}]{brink1993semiclassical}%
  \BibitemOpen
  \bibfield  {author} {\bibinfo {author} {\bibfnamefont {D.~M.}\ \bibnamefont {Brink}},\ }\href@noop {} {\emph {\bibinfo {title} {Semi-classical Methods for Nuclear Physics}}}\ (\bibinfo  {publisher} {Cambridge University Press},\ \bibinfo {year} {1993})\BibitemShut {NoStop}%
\bibitem [{\citenamefont {Nicholson}\ and\ \citenamefont {Nicholson}(1983)}]{nicholson1983introduction}%
  \BibitemOpen
  \bibfield  {author} {\bibinfo {author} {\bibfnamefont {D.~R.}\ \bibnamefont {Nicholson}}\ and\ \bibinfo {author} {\bibfnamefont {D.~R.}\ \bibnamefont {Nicholson}},\ }\href@noop {} {\emph {\bibinfo {title} {Introduction to plasma theory}}},\ Vol.~\bibinfo {volume} {1}\ (\bibinfo  {publisher} {Wiley New York},\ \bibinfo {year} {1983})\BibitemShut {NoStop}%
\bibitem [{\citenamefont {Sankaranarayanan}\ and\ \citenamefont {Lakshminarayan}(2003)}]{sankaranarayanan2003recurrence}%
  \BibitemOpen
  \bibfield  {author} {\bibinfo {author} {\bibfnamefont {R.}~\bibnamefont {Sankaranarayanan}}\ and\ \bibinfo {author} {\bibfnamefont {A.}~\bibnamefont {Lakshminarayan}},\ }\href@noop {} {\bibfield  {journal} {\bibinfo  {journal} {Phys. Rev. E}\ }\textbf {\bibinfo {volume} {68}},\ \bibinfo {pages} {036216} (\bibinfo {year} {2003})}\BibitemShut {NoStop}%
\bibitem [{\citenamefont {Anand}\ \emph {et~al.}(2024)\citenamefont {Anand}, \citenamefont {Davis},\ and\ \citenamefont {Ghose}}]{amit2024}%
  \BibitemOpen
  \bibfield  {author} {\bibinfo {author} {\bibfnamefont {A.}~\bibnamefont {Anand}}, \bibinfo {author} {\bibfnamefont {J.}~\bibnamefont {Davis}},\ and\ \bibinfo {author} {\bibfnamefont {S.}~\bibnamefont {Ghose}},\ }\href@noop {} {\bibfield  {journal} {\bibinfo  {journal} {Phys. Rev. Res.}\ }\textbf {\bibinfo {volume} {6}},\ \bibinfo {pages} {023120} (\bibinfo {year} {2024})}\BibitemShut {NoStop}%
\bibitem [{\citenamefont {Anand}\ \emph {et~al.}(2025)\citenamefont {Anand}, \citenamefont {Valluri}, \citenamefont {Davis},\ and\ \citenamefont {Ghose}}]{anand2025quantum}%
  \BibitemOpen
  \bibfield  {author} {\bibinfo {author} {\bibfnamefont {A.}~\bibnamefont {Anand}}, \bibinfo {author} {\bibfnamefont {D.}~\bibnamefont {Valluri}}, \bibinfo {author} {\bibfnamefont {J.}~\bibnamefont {Davis}},\ and\ \bibinfo {author} {\bibfnamefont {S.}~\bibnamefont {Ghose}},\ }\href@noop {} {\bibfield  {journal} {\bibinfo  {journal} {arXiv preprint arXiv:2508.09933}\ } (\bibinfo {year} {2025})}\BibitemShut {NoStop}%
\bibitem [{\citenamefont {Sharma}\ and\ \citenamefont {Bhosale}(2024{\natexlab{a}})}]{sharma2024exactly}%
  \BibitemOpen
  \bibfield  {author} {\bibinfo {author} {\bibfnamefont {H.}~\bibnamefont {Sharma}}\ and\ \bibinfo {author} {\bibfnamefont {U.~T.}\ \bibnamefont {Bhosale}},\ }\href@noop {} {\bibfield  {journal} {\bibinfo  {journal} {Phys. Rev. B}\ }\textbf {\bibinfo {volume} {109}},\ \bibinfo {pages} {014412} (\bibinfo {year} {2024}{\natexlab{a}})}\BibitemShut {NoStop}%
\bibitem [{\citenamefont {Sharma}\ and\ \citenamefont {Bhosale}(2024{\natexlab{b}})}]{sharma2024signatures}%
  \BibitemOpen
  \bibfield  {author} {\bibinfo {author} {\bibfnamefont {H.}~\bibnamefont {Sharma}}\ and\ \bibinfo {author} {\bibfnamefont {U.~T.}\ \bibnamefont {Bhosale}},\ }\href@noop {} {\bibfield  {journal} {\bibinfo  {journal} {Phys. Rev. B}\ }\textbf {\bibinfo {volume} {110}},\ \bibinfo {pages} {064313} (\bibinfo {year} {2024}{\natexlab{b}})}\BibitemShut {NoStop}%
\bibitem [{\citenamefont {Purohit}\ and\ \citenamefont {Bhosale}(2025)}]{purohit2025double}%
  \BibitemOpen
  \bibfield  {author} {\bibinfo {author} {\bibfnamefont {A.~V.}\ \bibnamefont {Purohit}}\ and\ \bibinfo {author} {\bibfnamefont {U.~T.}\ \bibnamefont {Bhosale}},\ }\href@noop {} {\bibfield  {journal} {\bibinfo  {journal} {Phys. Rev. E}\ }\textbf {\bibinfo {volume} {112}},\ \bibinfo {pages} {014217} (\bibinfo {year} {2025})}\BibitemShut {NoStop}%
\bibitem [{\citenamefont {Gorin}\ \emph {et~al.}(2006)\citenamefont {Gorin}, \citenamefont {Prosen}, \citenamefont {Seligman},\ and\ \citenamefont {{\v{Z}}nidari{\v{c}}}}]{gorin2006dynamics}%
  \BibitemOpen
  \bibfield  {author} {\bibinfo {author} {\bibfnamefont {T.}~\bibnamefont {Gorin}}, \bibinfo {author} {\bibfnamefont {T.}~\bibnamefont {Prosen}}, \bibinfo {author} {\bibfnamefont {T.~H.}\ \bibnamefont {Seligman}},\ and\ \bibinfo {author} {\bibfnamefont {M.}~\bibnamefont {{\v{Z}}nidari{\v{c}}}},\ }\href@noop {} {\bibfield  {journal} {\bibinfo  {journal} {Physics Reports}\ }\textbf {\bibinfo {volume} {435}},\ \bibinfo {pages} {33} (\bibinfo {year} {2006})}\BibitemShut {NoStop}%
\bibitem [{\citenamefont {Jalabert}\ and\ \citenamefont {Pastawski}(2001)}]{jalabert2001environment}%
  \BibitemOpen
  \bibfield  {author} {\bibinfo {author} {\bibfnamefont {R.~A.}\ \bibnamefont {Jalabert}}\ and\ \bibinfo {author} {\bibfnamefont {H.~M.}\ \bibnamefont {Pastawski}},\ }\href@noop {} {\bibfield  {journal} {\bibinfo  {journal} {Physical Rev. Lett.}\ }\textbf {\bibinfo {volume} {86}},\ \bibinfo {pages} {2490} (\bibinfo {year} {2001})}\BibitemShut {NoStop}%
\bibitem [{\citenamefont {Andraschko}\ and\ \citenamefont {Sirker}(2014)}]{andraschko2014dynamical}%
  \BibitemOpen
  \bibfield  {author} {\bibinfo {author} {\bibfnamefont {F.}~\bibnamefont {Andraschko}}\ and\ \bibinfo {author} {\bibfnamefont {J.}~\bibnamefont {Sirker}},\ }\href@noop {} {\bibfield  {journal} {\bibinfo  {journal} {Phys. Rev. B}\ }\textbf {\bibinfo {volume} {89}},\ \bibinfo {pages} {125120} (\bibinfo {year} {2014})}\BibitemShut {NoStop}%
\bibitem [{\citenamefont {Prosen}\ \emph {et~al.}(2003)\citenamefont {Prosen}, \citenamefont {Seligman},\ and\ \citenamefont {{\v{Z}}nidari{\v{c}}}}]{prosen2003theory}%
  \BibitemOpen
  \bibfield  {author} {\bibinfo {author} {\bibfnamefont {T.}~\bibnamefont {Prosen}}, \bibinfo {author} {\bibfnamefont {T.~H.}\ \bibnamefont {Seligman}},\ and\ \bibinfo {author} {\bibfnamefont {M.}~\bibnamefont {{\v{Z}}nidari{\v{c}}}},\ }\href@noop {} {\bibfield  {journal} {\bibinfo  {journal} {Progress of Theoretical Physics Supplement}\ }\textbf {\bibinfo {volume} {150}},\ \bibinfo {pages} {200} (\bibinfo {year} {2003})}\BibitemShut {NoStop}%
\bibitem [{\citenamefont {Yang}\ \emph {et~al.}(2019)\citenamefont {Yang}, \citenamefont {Zhou}, \citenamefont {Ma}, \citenamefont {Kong}, \citenamefont {Wang}, \citenamefont {Qin}, \citenamefont {Rong}, \citenamefont {Wang}, \citenamefont {Shi}, \citenamefont {Gong},\ and\ \citenamefont {Du}}]{Yang}%
  \BibitemOpen
  \bibfield  {author} {\bibinfo {author} {\bibfnamefont {K.}~\bibnamefont {Yang}}, \bibinfo {author} {\bibfnamefont {L.}~\bibnamefont {Zhou}}, \bibinfo {author} {\bibfnamefont {W.}~\bibnamefont {Ma}}, \bibinfo {author} {\bibfnamefont {X.}~\bibnamefont {Kong}}, \bibinfo {author} {\bibfnamefont {P.}~\bibnamefont {Wang}}, \bibinfo {author} {\bibfnamefont {X.}~\bibnamefont {Qin}}, \bibinfo {author} {\bibfnamefont {X.}~\bibnamefont {Rong}}, \bibinfo {author} {\bibfnamefont {Y.}~\bibnamefont {Wang}}, \bibinfo {author} {\bibfnamefont {F.}~\bibnamefont {Shi}}, \bibinfo {author} {\bibfnamefont {J.}~\bibnamefont {Gong}},\ and\ \bibinfo {author} {\bibfnamefont {J.}~\bibnamefont {Du}},\ }\href@noop {} {\bibfield  {journal} {\bibinfo  {journal} {Phys. Rev. B}\ }\textbf {\bibinfo {volume} {100}},\ \bibinfo {pages} {085308} (\bibinfo {year} {2019})}\BibitemShut {NoStop}%
\bibitem [{\citenamefont {Berdanier}\ \emph {et~al.}(2017)\citenamefont {Berdanier}, \citenamefont {Kolodrubetz}, \citenamefont {Vasseur},\ and\ \citenamefont {Moore}}]{berdanier2017floquet}%
  \BibitemOpen
  \bibfield  {author} {\bibinfo {author} {\bibfnamefont {W.}~\bibnamefont {Berdanier}}, \bibinfo {author} {\bibfnamefont {M.}~\bibnamefont {Kolodrubetz}}, \bibinfo {author} {\bibfnamefont {R.}~\bibnamefont {Vasseur}},\ and\ \bibinfo {author} {\bibfnamefont {J.~E.}\ \bibnamefont {Moore}},\ }\href@noop {} {\bibfield  {journal} {\bibinfo  {journal} {Phys. Rev. Lett.}\ }\textbf {\bibinfo {volume} {118}},\ \bibinfo {pages} {260602} (\bibinfo {year} {2017})}\BibitemShut {NoStop}%
\bibitem [{\citenamefont {Naji}\ \emph {et~al.}(2022)\citenamefont {Naji}, \citenamefont {Jafari}, \citenamefont {Zhou},\ and\ \citenamefont {Langari}}]{Naji}%
  \BibitemOpen
  \bibfield  {author} {\bibinfo {author} {\bibfnamefont {J.}~\bibnamefont {Naji}}, \bibinfo {author} {\bibfnamefont {R.}~\bibnamefont {Jafari}}, \bibinfo {author} {\bibfnamefont {L.}~\bibnamefont {Zhou}},\ and\ \bibinfo {author} {\bibfnamefont {A.}~\bibnamefont {Langari}},\ }\href@noop {} {\bibfield  {journal} {\bibinfo  {journal} {Phys. Rev. B}\ }\textbf {\bibinfo {volume} {106}},\ \bibinfo {pages} {094314} (\bibinfo {year} {2022})}\BibitemShut {NoStop}%
\bibitem [{\citenamefont {Jafari}\ \emph {et~al.}(2022)\citenamefont {Jafari}, \citenamefont {Akbari}, \citenamefont {Mishra},\ and\ \citenamefont {Johannesson}}]{Jafari}%
  \BibitemOpen
  \bibfield  {author} {\bibinfo {author} {\bibfnamefont {R.}~\bibnamefont {Jafari}}, \bibinfo {author} {\bibfnamefont {A.}~\bibnamefont {Akbari}}, \bibinfo {author} {\bibfnamefont {U.}~\bibnamefont {Mishra}},\ and\ \bibinfo {author} {\bibfnamefont {H.}~\bibnamefont {Johannesson}},\ }\href@noop {} {\bibfield  {journal} {\bibinfo  {journal} {Phys. Rev. B}\ }\textbf {\bibinfo {volume} {105}},\ \bibinfo {pages} {094311} (\bibinfo {year} {2022})}\BibitemShut {NoStop}%
\bibitem [{sup()}]{supplementary2025}%
  \BibitemOpen
  \href@noop {} {\bibinfo  {journal} {This supplementary material [] provides detailed analytical calculations the DKT with $k_r = j\pi/2$ for even and odd qubits. We also give derivation for even qubits case of the DKT with $k_r = j\pi$, along with numerical results of Husimi distributions supporting the main text.}\ }\BibitemShut {NoStop}%
\bibitem [{\citenamefont {Agarwal}(1981)}]{agarwal1981relation}%
  \BibitemOpen
\bibfield  {journal} {  }\bibfield  {author} {\bibinfo {author} {\bibfnamefont {G.~S.}\ \bibnamefont {Agarwal}},\ }\href@noop {} {\bibfield  {journal} {\bibinfo  {journal} {Phys. Rev. A}\ }\textbf {\bibinfo {volume} {24}},\ \bibinfo {pages} {2889} (\bibinfo {year} {1981})}\BibitemShut {NoStop}%
\bibitem [{\citenamefont {Zarum}\ and\ \citenamefont {Sarkar}(1998)}]{Zarum1998}%
  \BibitemOpen
  \bibfield  {author} {\bibinfo {author} {\bibfnamefont {R.}~\bibnamefont {Zarum}}\ and\ \bibinfo {author} {\bibfnamefont {S.}~\bibnamefont {Sarkar}},\ }\href@noop {} {\bibfield  {journal} {\bibinfo  {journal} {Phys. Rev. E}\ }\textbf {\bibinfo {volume} {57}},\ \bibinfo {pages} {5467} (\bibinfo {year} {1998})}\BibitemShut {NoStop}%
\bibitem [{\citenamefont {Heyl}\ \emph {et~al.}(2013)\citenamefont {Heyl}, \citenamefont {Polkovnikov},\ and\ \citenamefont {Kehrein}}]{Heyl2013dynamical}%
  \BibitemOpen
  \bibfield  {author} {\bibinfo {author} {\bibfnamefont {M.}~\bibnamefont {Heyl}}, \bibinfo {author} {\bibfnamefont {A.}~\bibnamefont {Polkovnikov}},\ and\ \bibinfo {author} {\bibfnamefont {S.}~\bibnamefont {Kehrein}},\ }\href@noop {} {\bibfield  {journal} {\bibinfo  {journal} {Phys. Rev. Lett.}\ }\textbf {\bibinfo {volume} {110}},\ \bibinfo {pages} {135704} (\bibinfo {year} {2013})}\BibitemShut {NoStop}%
\bibitem [{\citenamefont {Gritsev}\ and\ \citenamefont {Polkovnikov}(2017)}]{gritsev2017integrable}%
  \BibitemOpen
  \bibfield  {author} {\bibinfo {author} {\bibfnamefont {V.}~\bibnamefont {Gritsev}}\ and\ \bibinfo {author} {\bibfnamefont {A.}~\bibnamefont {Polkovnikov}},\ }\href@noop {} {\bibfield  {journal} {\bibinfo  {journal} {SciPost Physics}\ }\textbf {\bibinfo {volume} {2}},\ \bibinfo {pages} {021} (\bibinfo {year} {2017})}\BibitemShut {NoStop}%
\bibitem [{\citenamefont {Mehta}(2004)}]{mehta2004random}%
  \BibitemOpen
  \bibfield  {author} {\bibinfo {author} {\bibfnamefont {M.~L.}\ \bibnamefont {Mehta}},\ }\href@noop {} {\emph {\bibinfo {title} {Random Matrices}}},\ \bibinfo {edition} {3rd}\ ed.\ (\bibinfo  {publisher} {Elsevier Academic Press},\ \bibinfo {address} {London},\ \bibinfo {year} {2004})\BibitemShut {NoStop}%
\bibitem [{\citenamefont {Tekur}\ and\ \citenamefont {Santhanam}(2020)}]{harshini2020symmetry}%
  \BibitemOpen
  \bibfield  {author} {\bibinfo {author} {\bibfnamefont {S.~H.}\ \bibnamefont {Tekur}}\ and\ \bibinfo {author} {\bibfnamefont {M.~S.}\ \bibnamefont {Santhanam}},\ }\href@noop {} {\bibfield  {journal} {\bibinfo  {journal} {Phys. Rev. Res.}\ }\textbf {\bibinfo {volume} {2}},\ \bibinfo {pages} {032063} (\bibinfo {year} {2020})}\BibitemShut {NoStop}%
\bibitem [{\citenamefont {Tekur}\ \emph {et~al.}(2018)\citenamefont {Tekur}, \citenamefont {Bhosale},\ and\ \citenamefont {Santhanam}}]{bhosale2028higher}%
  \BibitemOpen
  \bibfield  {author} {\bibinfo {author} {\bibfnamefont {S.~H.}\ \bibnamefont {Tekur}}, \bibinfo {author} {\bibfnamefont {U.~T.}\ \bibnamefont {Bhosale}},\ and\ \bibinfo {author} {\bibfnamefont {M.~S.}\ \bibnamefont {Santhanam}},\ }\href@noop {} {\bibfield  {journal} {\bibinfo  {journal} {Phys. Rev. B}\ }\textbf {\bibinfo {volume} {98}},\ \bibinfo {pages} {104305} (\bibinfo {year} {2018})}\BibitemShut {NoStop}%
\bibitem [{\citenamefont {Meier}\ \emph {et~al.}(2019)\citenamefont {Meier}, \citenamefont {Ang'ong'a}, \citenamefont {An},\ and\ \citenamefont {Gadway}}]{meier2019exploring}%
  \BibitemOpen
  \bibfield  {author} {\bibinfo {author} {\bibfnamefont {E.~J.}\ \bibnamefont {Meier}}, \bibinfo {author} {\bibfnamefont {J.}~\bibnamefont {Ang'ong'a}}, \bibinfo {author} {\bibfnamefont {F.~A.}\ \bibnamefont {An}},\ and\ \bibinfo {author} {\bibfnamefont {B.}~\bibnamefont {Gadway}},\ }\href@noop {} {\bibfield  {journal} {\bibinfo  {journal} {Phys. Rev. A}\ }\textbf {\bibinfo {volume} {100}},\ \bibinfo {pages} {013623} (\bibinfo {year} {2019})}\BibitemShut {NoStop}%
\end{thebibliography}%


\makeatletter
\let\addcontentsline\addcontentslineOriginal
\makeatother

\renewcommand{\thesection}{S\Roman{section}}
\renewcommand{\thesubsection}{\Alph{subsection}}
\renewcommand{\theequation}{S\arabic{equation}}
\renewcommand{\thefigure}{S\arabic{figure}}
\renewcommand{\thetable}{S\arabic{table}}

\setcounter{section}{0}
\setcounter{subsection}{0}
\setcounter{equation}{0}
\setcounter{figure}{0}
\setcounter{table}{0}

\renewcommand{\theHequation}{S\arabic{equation}}
\renewcommand{\theHfigure}{S\arabic{figure}}
\renewcommand{\theHtable}{S\arabic{table}}


\newcommand{\suppsection}[2][]{%
  \refstepcounter{section}%
  \phantomsection%
  \section*{\texorpdfstring{S\Roman{section}\quad #2}{S\Roman{section} #2}}%
  \addcontentsline{toc}{section}{S\Roman{section}\quad #2}%
  \ifstrempty{#1}{}{%
    \label{#1}%
  }%
}

\newcommand{\suppsubsection}[2][]{%
  \refstepcounter{subsection}%
  \phantomsection%
  \subsection*{#2}
  \addcontentsline{toc}{subsection}{\thesubsection\quad #2}%
  \ifstrempty{#1}{}{%
    \label{#1}%
  }%
}

\clearpage
\onecolumngrid

\begin{center}
\textbf{\large Supplementary Material for\\ 
``\textit{Floquet Recurrences in the Double Kicked Top}''}
\end{center}

\tableofcontents

\suppsection{Introduction}
This Supplementary Material presents detailed analytical calculations supporting the periodicity of the Floquet operator introduced in the main text. In Sec.~\ref{supsec:jpiby2}, we derive the periodicity for both even and odd qubit systems at $k_r = j\pi/2$, while Sec.~\ref{supsec:jpiby4} addresses the case $k_r = j\pi/4$.

\suppsection{Transformed kick strength \texorpdfstring{$k_r = j \pi/2$}{}}\label{supsec:jpiby2}
In this section we derive periodicity of the Floquet operator of the double kicked top for integer and half-odd integer values of $j$. The Floquet operator is given by 
\begin{equation}\label{Floquet}
    \mathcal{U} = \exp\left[- i \left(\frac{k_r - k_\theta}{2j}\right)J_x^2\right]\exp\left[- i \left(\frac{k_r + k_\theta}{2j} \right)J_z^2\right]\exp\left(- i \frac{\pi}{2} J_y\right).
\end{equation}
We show that the above Floquet operator is periodic for $k_r = j\pi/2$ and independent of $k_\theta$. Denoting $\sigma_x^{(m)}$, $\sigma_z^{(m)}$ and $\sigma_y^{(m)}$ as Pauli operators acting on $m$-th qubit \citep{amit2024}, we re-write the above Floquet operator as follows:
\begin{align}\label{eq3}
    \mathcal{U} = \exp\left[- i \left(\frac{k_r - k_\theta}{8j}\right) {\left(\sum_{a=1}^{2j} \sigma_x^{(a)}\right)}^2 \right] \exp\left[- i \left(\frac{k_r + k_\theta}{8j}\right) {\left(\sum_{a=1}^{2j} \sigma_y^{(a)} \right)}^2 \right] \exp\left(- i \frac{\pi}{4} \sum_{a=1}^{2j} \sigma_y^{(a)}\right).
\end{align}
Using identity ${\left(\sum_{a=1}^{2j} \sigma_x^{(a)}\right)}^2 = \sum_{a=1}^{2j} \mathds{I} + 2 \sum_{a < b = 2}^{2j} \sigma_x^{(a)} \sigma_x^{(b)} $, we get
\begin{equation}
    \exp\left[- i \left(\frac{k_r - k_\theta}{8j}\right) {\left(\sum_{a=1}^{2j} \sigma_x^{(a)}\right)}^2 \right]  = \exp\left[- i \left(\frac{k_r - k_\theta}{8j}\right) \mathds{I}^{\otimes 2j} \right] \exp\left[- i \left(\frac{k_r - k_\theta}{4j}\right) \sum_{a < b = 2}^{2j} \sigma_x^{(a)} \sigma_x^{(b)} \right]. 
\end{equation}
Similarly, for the second operator, we get
\begin{equation}
    \exp\left[- i \left(\frac{k_r + k_\theta}{8j}\right) {\left(\sum_{a=1}^{2j} \sigma_z^{(a)}\right)}^2 \right]  = \exp\left[- i \left(\frac{k_r + k_\theta}{8j}\right) \mathds{I}^{\otimes 2j} \right] \exp\left[- i \left(\frac{k_r + k_\theta}{4j}\right) \sum_{a < b = 2}^{2j} \sigma_z^{(a)} \sigma_z^{(b)} \right]. 
\end{equation}
Thus, the non-local part of the Floquet operator becomes
\begin{align}
    \exp\left[- i \frac{k_r - k_\theta}{8j} {\left(\sum_{a=1}^{2j} \sigma_x^{(a)}\right)}^2 \right] \exp\left[- i \frac{k_r + k_\theta}{8j} {\left(\sum_{a=1}^{2j} \sigma_y^{(a)}\right)}^2\right] =& \; \exp\left(- i \frac{k_r}{4j} 2j\right) \exp\left[i \frac{k_r - k_\theta}{4j} \sum_{a < b = 2}^{2j} {(i \sigma_x)}^{(a)} {(i \sigma_x)}^{(b)} \right] \notag \\
    &\; \times \exp\left[ i \frac{k_r + k_\theta}{4j} \sum_{a < b = 2}^{2j} {(i \sigma_z)}^{(a)} {(i \sigma_z)}^{(b)}\right]. 
\end{align}

\suppsubsection{Integer \texorpdfstring{$j$}{}}
For integer values of $j$, the non-local part of the Floquet operator simplifies as follows:
\begin{align}
    \exp &\left[- i \frac{k_r - k_\theta}{8j} {\left(\sum_{a=1}^{2j} \sigma_x^{(a)}\right)}^2 \right] \exp\left[- i \frac{k_r + k_\theta}{8j} {\left(\sum_{a=1}^{2j} \sigma_z^{(a)}\right)}^2\right] \notag \\
    \quad \quad &=  \exp\left(- i \frac{k_r}{4j} 2j\right) \prod_{a < b = 2}^{2j} \exp\left[ i \frac{k_r - k_\theta}{4j} {(i \sigma_x)}^{(a)} {(i \sigma_x)}^{(b)} \right] \prod_{c < d = 2}^{2j} \exp \left[ i \frac{k_r + k_\theta}{4j} {(i \sigma_z)}^{(c)} {(i \sigma_z)}^{(d)} \right] \notag \\
    \quad \quad &=  \exp\left(- i \frac{k_r}{4j} 2j\right) \left[\mathds{I}^{\otimes 2j} \cos\left(\frac{k_r - k_\theta}{4j}\right) + i {(i\sigma_x)}^{\otimes 2j} \sin\left(\frac{k_r - k_\theta}{4j}\right)\right] \left[\mathds{I}^{\otimes 2j} \cos\left(\frac{k_r + k_\theta}{4j}\right) + i {(i\sigma_z)}^{\otimes 2j} \sin\left(\frac{k_r + k_\theta}{4j}\right)\right].
\end{align}
Now, by setting $k_r = j\pi/2$ the Floquet operator gets further simplified as follows:
\begin{align}
    \mathcal{U} =& \; \frac{e^{-i\frac{\pi}{8}2j}}{2} \left\lbrace \left[\mathds{I}^{\otimes 2j} + {(i\sigma_y)}^{\otimes 2j}\right]\cos\left(\frac{\pi}{4}\right) + i \left[{(i\sigma_z)}^{\otimes 2j} + {(i\sigma_x)}^{\otimes 2j}\right]\sin\left(\frac{\pi}{4}\right) \right. \notag \\
    &\;\left. + \left[\mathds{I}^{\otimes 2j} - {(i\sigma_y)}^{\otimes 2j}\right] \cos\left(\frac{k_\theta}{2j}\right) + i \left[{(i\sigma_z)}^{\otimes 2j} - {(i\sigma_x)}^{\otimes 2j}\right]\sin\left(\frac{k_\theta}{2j}\right) \right\rbrace \gamma^{\otimes 2j}.
\end{align}
Here, $\gamma = e^{-i\frac{\pi}{4}\sigma_y}$. Using relations $\gamma \sigma_z = \sigma_x \gamma$ and $\sigma_z \gamma = -\gamma \sigma_x$ \citep{amit2024}, we show that $\mathcal{U}^2$ does not depend on $k_\theta$. It is given by
\begin{align}
    \mathcal{U}^2 = \; \frac{e^{-i\frac{\pi}{4}2j}}{4} \left[ C_1 + C_2 \cos\left(\frac{k_\theta}{2j}\right) + C_3\sin\left(\frac{k_\theta}{2j}\right) \right] \left[ C_1 + C_2 \cos\left(\frac{k_\theta}{2j}\right) - C_3\sin\left(\frac{k_\theta}{2j}\right) \right] {\left(\gamma^{\otimes 2j}\right)}^2,
\end{align}
where,
\begin{align}
    C_1 = \left[\mathds{I}^{\otimes 2j} + {(i\sigma_y)}^{\otimes 2j}\right]\cos\left(\frac{\pi}{4}\right) + i \left[{(i\sigma_z)}^{\otimes 2j} + {(i\sigma_x)}^{\otimes 2j}\right]\sin\left(\frac{\pi}{4}\right), C_2 = \mathds{I}^{\otimes 2j} - {(i\sigma_y)}^{\otimes 2j} \text{ and }\; C_3 = i \left[{(i\sigma_z)}^{\otimes 2j} - {(i\sigma_x)}^{\otimes 2j}\right].
\end{align}
For the case of even-$2j$, matrices ${(i\sigma_x)}^{\otimes 2j}$, ${(i\sigma_y)}^{\otimes 2j}$ and ${(i\sigma_z)}^{\otimes 2j}$ commute with each other. Further, ${(i\sigma_x)}^{\otimes 2j}{(i\sigma_y)}^{\otimes 2j} = {(-i\sigma_z)}^{\otimes 2j}$, ${(i\sigma_y)}^{\otimes 2j}{(i\sigma_z)}^{\otimes 2j} = {(-i\sigma_x)}^{\otimes 2j}$ and ${(i\sigma_y)}^{\otimes 2j}{(i\sigma_z)}^{\otimes 2j} = {(-i\sigma_x)}^{\otimes 2j}$. Using these properties, we get
\begin{align}\label{eq:pauli_identities}
    \left[\mathds{I}^{\otimes 2j} \pm {(i\sigma_y)}^{\otimes 2j}, {(i\sigma_z)}^{\otimes 2j} - {(i\sigma_x)}^{\otimes 2j}\right] &= 0, \quad C_3^2 = - 2 \left(\mathds{I}^{\otimes 2j} - {(i\sigma_y)}^{\otimes 2j}\right), \quad C_2^2 = 2\left(\mathds{I}^{\otimes 2j} - {(i\sigma_y)}^{\otimes 2j}\right), \notag \\
    \left\lbrace \mathds{I}^{\otimes 2j} + {(i\sigma_y)}^{\otimes 2j}, \mathds{I}^{\otimes 2j} - {(i\sigma_y)}^{\otimes 2j}\right\rbrace &= 0 \quad \text{and}\quad \left\lbrace {(i\sigma_z)}^{\otimes 2j} + {(i\sigma_x)}^{\otimes 2j} , \mathds{I}^{\otimes 2j} - {(i\sigma_y)}^{\otimes 2j}\right\rbrace = 0
\end{align}
Then, the coefficients $C_i$'s satisfy $\lbrace C_1, C_2\rbrace = 0$, $\left[C_3, C_1\right] = 0$, $\left[ C_3, C_2\right] = 0$ and $C_3^2 = - C_2^2$. As a result,
\begin{align}
    \mathcal{U}^2 = \; \frac{e^{-i\frac{\pi}{4}2j}}{4} \left( C_1^2 + C_2^2 \right) {\left(\gamma^{\otimes 2j}\right)}^2 = \frac{e^{-i\frac{\pi}{4}2j}}{2} \left( \mathds{I}^{\otimes 2j} - {(i\sigma_y)}^{\otimes 2j} + i{(i\sigma_z)}^{\otimes 2j} + i{(i\sigma_x)}^{\otimes 2j}\right) {\left(\gamma^{\otimes 2j}\right)}^2,
\end{align}
becomes independent of $k_\theta$. Finally, the forth power of the Floquet operator is obtained as follows: 
\begin{align}
    \mathcal{U}^4 = - e^{i\frac{\pi}{2}2j} {(i\sigma_y)}^{\otimes 2j} \implies \mathcal{U}^8 = - \mathds{I}^{\otimes 2j}.
\end{align}
This result is used in the main text.

\suppsubsection{Half-odd integer \texorpdfstring{$j$}{}}\label{subsec:2}
In this subsection, we show that $\mathcal{U}^3$ is independent of $k_\theta$ and find the period. We simplify the operator $\mathcal{U}^2$ by swapping $\gamma^{\otimes 2j}$ to the right side, using $J_z^2 \to J_x^2$ and $J_x^2 \to J_z^2$ as follows:
\begin{align}
    \mathcal{U}^2 =& \; \exp\left(- i \frac{k'}{2j}J_x^2\right) \exp\left(- i \frac{k}{2j}J_z^2\right)  \exp\left(- i \frac{k'}{2j}J_z^2\right) \exp\left(- i \frac{k}{2j}J_x^2\right) {\left(\gamma^{\otimes 2j}\right)}^2 \notag \\
    =& \; \exp\left(- i \frac{k'}{2j}J_x^2\right) \exp\left(- i \frac{k_r}{j}J_z^2\right)  \exp\left(- i \frac{k}{2j}J_x^2\right) {\left(\gamma^{\otimes 2j}\right)}^2.
\end{align}
Since identities \eqref{eq:pauli_identities} do not hold for the case of odd-$2j$, the above operator does not get simplified. Therefore, we proceed to find $\mathcal{U}^3 = \mathcal{U}\cdot \mathcal{U}^2$ as follows:
\begin{align}\label{eq:jpiby2_odd}
    \mathcal{U}\cdot \mathcal{U}^2 =& \; \exp\left(- i \frac{k'}{2j}J_x^2\right) \exp\left(- i \frac{k}{2j}J_z^2\right) \gamma^{\otimes 2j} \exp\left(- i \frac{k'}{2j}J_x^2\right) \exp\left(- i \frac{k_r}{j}J_z^2\right)  \exp\left(- i \frac{k}{2j}J_x^2\right) {\left(\gamma^{\otimes 2j}\right)}^2 \notag \\
    =& \; \exp\left(- i \frac{k'}{2j}J_x^2\right) \exp\left(- i \frac{k}{2j}J_z^2\right) \exp\left(- i \frac{k'}{2j}J_z^2\right) \exp\left(- i \frac{k_r}{j}J_x^2\right)  \exp\left(- i \frac{k}{2j}J_z^2\right) {\left(\gamma^{\otimes 2j}\right)}^3 \notag \\
    =& \; \exp\left(- i \frac{k'}{2j}J_x^2\right) \exp\left(- i \frac{k_r}{j}J_z^2\right) \exp\left(- i \frac{k_r}{j}J_x^2\right) \exp\left(- i \frac{k}{2j}J_z^2\right) {\left(\gamma^{\otimes 2j}\right)}^3.
\end{align}
Here, we have used the relation $k_r = (k + k')/2$. Now, using $k = k_r + k_\theta$, $k' = k_r - k_\theta$ and setting $k_r = j\pi/2$, we get
\begin{align}
    \mathcal{U}^3 = \exp\left(- i \frac{\pi}{4}J_x^2\right) \left[\exp\left(i \frac{k_\theta}{2j}J_x^2\right) \exp\left(- i \frac{\pi}{2}J_z^2\right) \exp\left(- i \frac{\pi}{2}J_x^2\right) \exp\left(- i \frac{k_\theta}{2j}J_z^2\right)\right] \exp\left(- i \frac{\pi}{4}J_z^2\right) {\left(\gamma^{\otimes 2j}\right)}^3.
\end{align}
The square bracket in the operator $ \mathcal{U}^3$ can be simplified using following identities:
\begin{align}
    \exp \left(i\frac{k_\theta}{4j}\sigma_x\right) \exp \left(-i\frac{\pi}{4}\sigma_z\right) &= \exp \left(-i\frac{\pi}{4}\sigma_z\right) \exp \left(-i\frac{k_\theta}{4j}\sigma_y\right) \text{ and} \notag \\
    \exp \left(-i\frac{\pi}{4}\sigma_x\right) \exp \left(-i\frac{k_\theta}{4j}\sigma_z\right) &= \exp \left(i\frac{k_\theta}{4j}\sigma_y\right) \exp \left(-i\frac{\pi}{4}\sigma_x\right).
\end{align}
For the case of odd-$2j$, we get 
\begin{align}
    \exp\left(i \frac{k_\theta}{2j}J_x^2\right) \exp\left(- i \frac{\pi}{2}J_z^2\right) =& \; \exp\left( i \frac{k_\theta}{8j}2j \right) \exp\left( -i \frac{\pi}{8}2j\right) \prod_{a < b = 2}^{2j} \exp\left( i \frac{k_\theta}{4j} \sigma_x^{(a)}\sigma_x^{(b)} \right) \prod_{c < d = 2}^{2j} \exp\left( -i \frac{\pi}{4} \sigma_z^{(c)}\sigma_z^{(d)} \right) \notag \\
    =& \; \exp\left( i \frac{k_\theta}{8j}2j\right) \exp\left( -i \frac{\pi}{8}2j\right) \prod_{a < b = 2}^{2j} \exp\left( -i \frac{\pi}{4} \sigma_z^{(a)}\sigma_z^{(a)} \right) \prod_{c < d = 2}^{2j} \exp\left( i \frac{k_\theta}{4j}  \sigma_y^{(c)}\sigma_y^{(d)} \right), \\
    \exp\left(- i \frac{\pi}{2}J_x^2\right) \exp\left(-i \frac{k_\theta}{2j}J_z^2\right) =& \; \exp\left( -i \frac{\pi}{8}2j\right) \exp\left( -i \frac{k_\theta}{8j}2j\right) \prod_{a < b = 2}^{2j} \exp\left( -i \frac{\pi}{4} \sigma_x^{(a)}\sigma_x^{(b)}\right) \prod_{c < d = 2}^{2j} \exp\left( -i \frac{k_\theta}{4j} \sigma_z^{(c)}\sigma_z^{(d)}\right) \notag \\
    =& \; \exp\left( -i \frac{\pi}{8}2j\right) \exp\left( -i \frac{k_\theta}{8j}2j\right) \prod_{a < b = 2}^{2j} \exp\left(  -i \frac{k_\theta}{4j}  \sigma_y^{(a)}\sigma_y^{(b)} \right) \prod_{c < d = 2}^{2j} \exp\left( -i \frac{\pi}{4} \sigma_x^{(c)}\sigma_x^{(d)}\right).
\end{align}
Combining these terms, we get the square bracketed terms as follows:
\begin{align}
    \exp\left(i \frac{k_\theta}{2j}J_x^2\right) \exp\left(- i \frac{\pi}{2}J_z^2\right) \exp\left(- i \frac{\pi}{2}J_x^2\right) \exp\left(- i \frac{k_\theta}{2j}J_z^2\right) = \exp\left(- i \frac{\pi}{2}J_z^2\right) \exp\left(- i \frac{\pi}{2}J_x^2\right).
\end{align}
Then, the floquet operator is given by
\begin{align}
    \mathcal{U}^3 = \exp\left(- i \frac{\pi}{4}J_x^2\right) \exp\left(- i \frac{\pi}{2}J_z^2\right) \exp\left(- i \frac{\pi}{2}J_x^2\right) \exp\left(- i \frac{\pi}{4}J_z^2\right) \cdot {\left(\gamma^{\otimes 2j}\right)}^3.
\end{align}
The sixth power of the Floquet operator is obtained by swapping ${\left(\gamma^{\otimes 2j}\right)}^3$ using properties of Pauli matrices, we get 
\begin{align}
    \mathcal{U}^6 &= \; \exp\left(- i \frac{\pi}{4}J_x^2\right) {\left[\exp\left(- i \frac{\pi}{2}J_z^2\right) \exp\left(- i \frac{\pi}{2}J_x^2\right)\right]}^2 \exp\left(- i \frac{\pi}{2}J_z^2\right) \exp\left(- i \frac{\pi}{4}J_x^2\right) \cdot {\left(\gamma^{\otimes 2j}\right)}^6 \notag \\
    &= \; \exp\left(- i \frac{\pi}{4}J_x^2\right) {\left[\exp\left(- i \frac{\pi}{2}J_z^2\right) \exp\left(- i \frac{\pi}{2}J_x^2\right)\right]}^3  \exp\left( i \frac{\pi}{4}J_x^2\right) \cdot {\left(\gamma^{\otimes 2j}\right)}^6.
\end{align}
The case of odd-$2j$ further satisfies the following identity: (see (B31) of Ref.~\citep{amit2024})
\begin{align}
    {\left[\exp\left( \pm i \frac{\pi}{2}J_a^2\right) \exp\left( \pm i \frac{\pi}{2}J_b^2\right)\right]}^3 = - \mathds{I}^{\otimes 2j} \quad \text{for }\;\; a, b \in \lbrace x,y,z\rbrace.
\end{align}
As a result, we get
\begin{align}
    \mathcal{U}^6 = - {(-1)}^{2j} \; {(i\sigma_y)}^{\otimes 2j} \implies \mathcal{U}^{12} = {(-1)}^{2j} \mathds{I}^{\otimes 2j}.
\end{align}
This result is used in the main text.

\suppsection{Transformed kick strength \texorpdfstring{$k_r = j \pi/4$}{}}\label{supsec:jpiby4}
In this section we study the Floquet operator of the double kicked top for $k_r = j\pi/4$ and show its independence of $k_\theta$. For integer values of $j$, the Floquet operator shows periodicity. The derivation remains same for both the cases, even-$2j$ and odd-$2j$ till Eq.~\eqref{eq:jpiby4_integer}.

\begin{figure}
    \includegraphics[width=\linewidth]{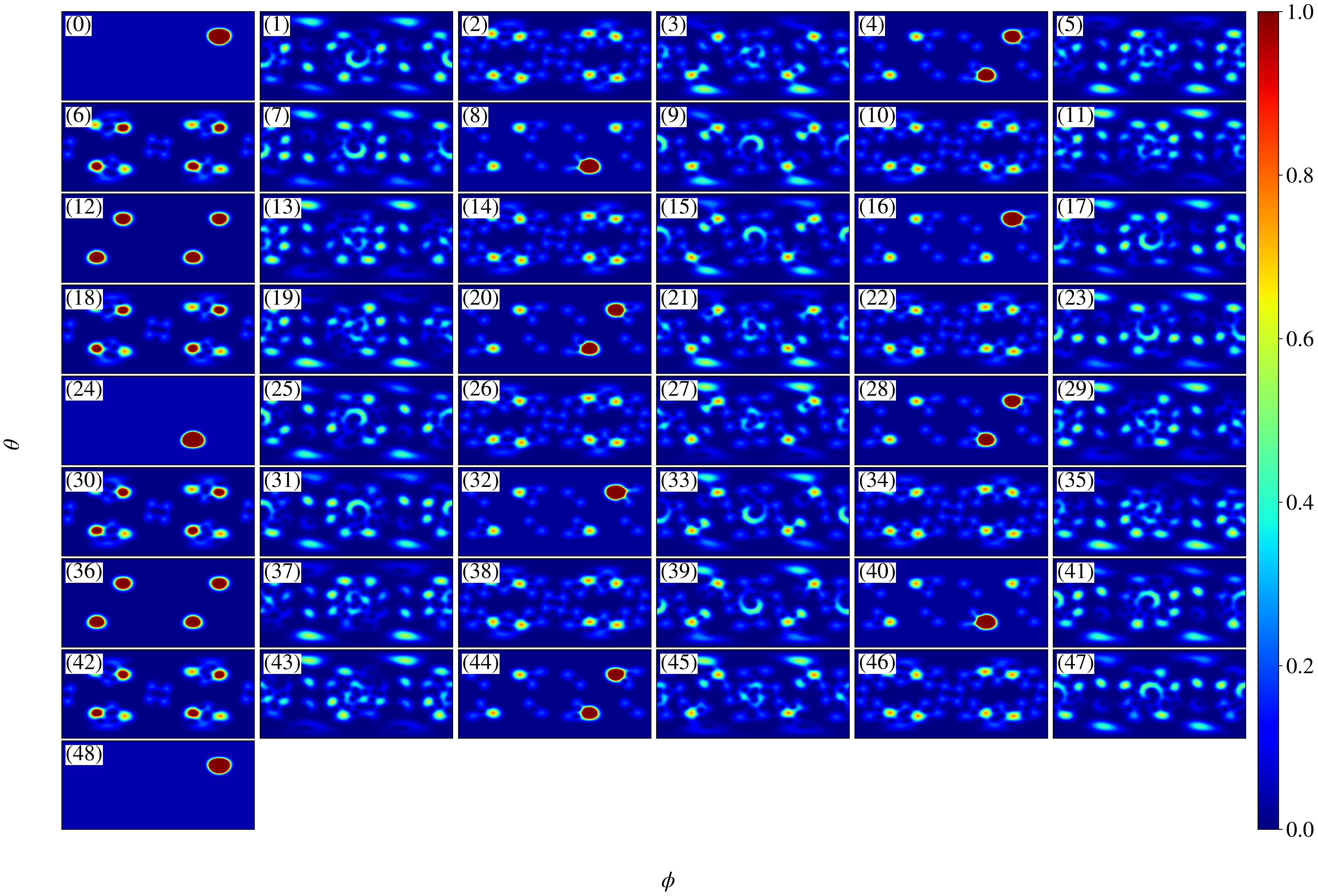}
    \caption{Husimi function of the time-evolved initial state $|\theta_0 = 2.25,\, \phi_0 = 2.0\rangle$ over eight kicks. Here, $k_r = j\pi/4$, $k_\theta = 0$, and $j = 76$. Panels (0)–(48) correspond to $n = 0$ through $n = 48$, respectively.}\label{suppfig:husimi_even_jpiby4}
\end{figure}

\begin{figure}
    \includegraphics[width=\linewidth]{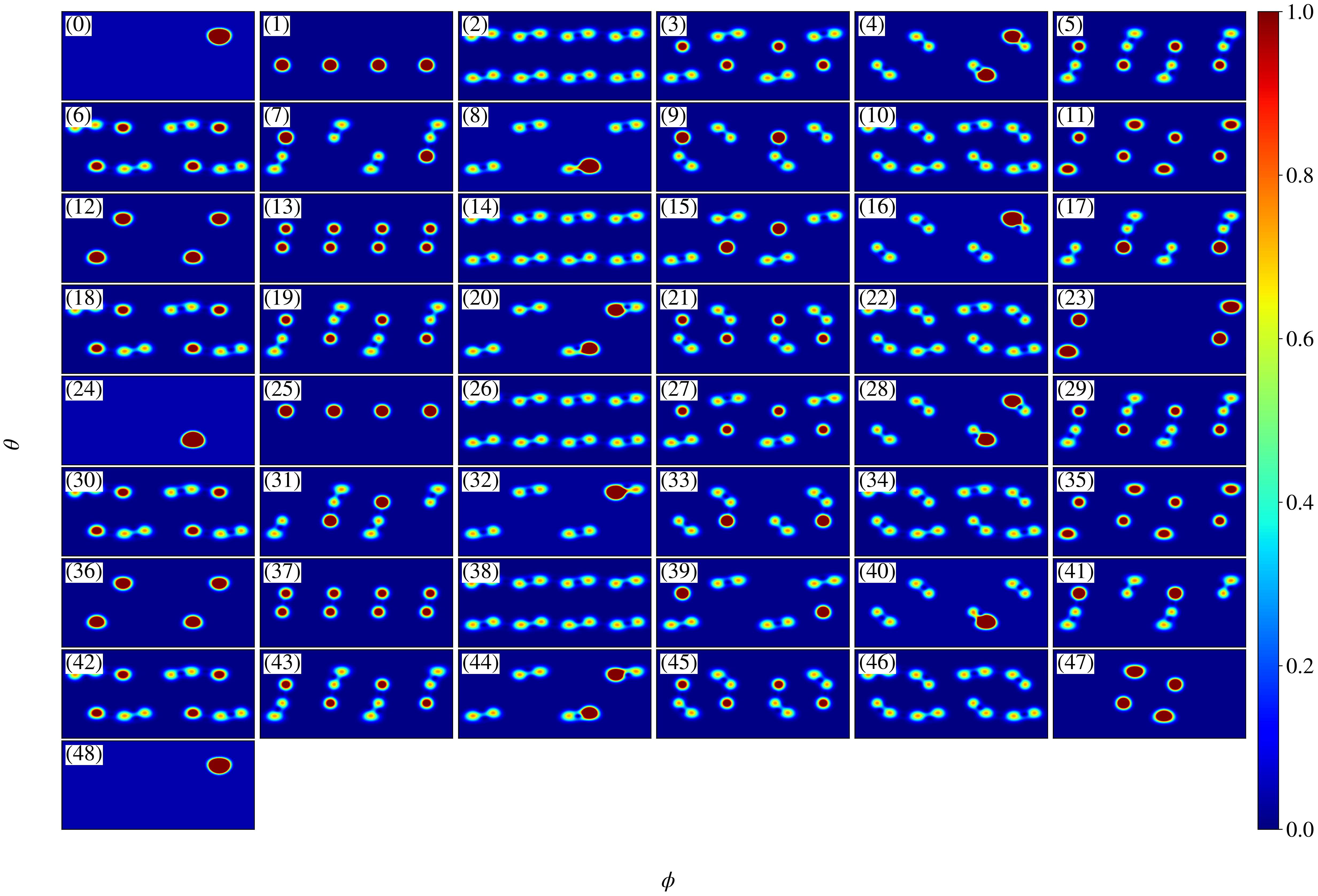}
    \caption{Husimi function of the time-evolved initial state $|\theta_0 = 2.25,\, \phi_0 = 2.0\rangle$ over eight kicks. Here, $k_r = j\pi/4$, $k_\theta = k_r$, and $j = 76$. Panels (0)–(48) correspond to $n = 0$ through $n = 48$, respectively.}\label{suppfig:husimi_even_qkt_jpiby4}
\end{figure}

\suppsubsection{Integer \texorpdfstring{$j$}{}}
Following the similar procedure discussed in the earlier section, we get the Floquet operator,
\begin{align}
    \mathcal{U}^3 = \; \exp\left(- i \frac{k'}{2j}J_x^2\right) \exp\left(- i \frac{\pi}{4}J_z^2\right) \exp\left(- i \frac{\pi}{4}J_x^2\right) \exp\left(- i \frac{k}{2j}J_z^2\right) \cdot {\left(\gamma^{\otimes 2j}\right)}^3.
\end{align}
Again taking ${\left(\gamma^{\otimes 2j}\right)}^3$ to the right by using properties of Pauli matrices, we get
\begin{align}
    \mathcal{U}^6 &= \; \exp\left(- i \frac{k'}{2j}J_x^2\right) {\left[\exp\left(- i \frac{\pi}{4}J_z^2\right) \exp\left(- i \frac{\pi}{4}J_x^2\right)\right]}^2 \exp\left(- i \frac{\pi}{4}J_z^2\right) \exp\left(- i \frac{k}{2j}J_x^2\right) \cdot {\left(\gamma^{\otimes 2j}\right)}^6 \notag \\
    &= \; \exp\left(- i \frac{k'}{2j}J_x^2\right) {\left[\exp\left(- i \frac{\pi}{4}J_z^2\right) \exp\left(- i \frac{\pi}{4}J_x^2\right)\right]}^3 \exp\left(i \frac{\pi}{4}J_x^2\right) \exp\left(- i \frac{k}{2j}J_x^2\right) \cdot {\left(\gamma^{\otimes 2j}\right)}^6.
\end{align}
In the second step, we have used properties $J_z^2 \to J_x^2$ and $J_x^2 \to J_z^2$ while swapping them with the operator ${\left(\gamma^{\otimes 2j}\right)}^3$. Proceeding further by squaring the above operator to get $\mathcal{U}^{12}$. Here, due to the even powers of $\gamma$, operators $J_z^2 \to J_z^2$ and $J_x^2 \to J_x^2$ remain unchanged while swapping ${\left(\gamma^{\otimes 2j}\right)}^6$ to the right. Then, simplifying the algebra, we get
\begin{align}
    \mathcal{U}^{12} = \; \exp\left(- i \frac{k'}{2j}J_x^2\right) {\left[\exp\left(- i \frac{\pi}{4}J_z^2\right) \exp\left(- i \frac{\pi}{4}J_x^2\right)\right]}^6 \exp\left(- i \frac{k}{2j}J_x^2\right) \;\exp\left(i \frac{\pi}{4}J_x^2\right) \cdot {\left(\gamma^{\otimes 2j}\right)}^{12}.
\end{align}
The square bracketed term can be written in qubit-form as follows:
\begin{align}\label{eq:jpiby4_integer}
    \exp\left(- i \frac{\pi}{4}J_z^2\right) \exp\left(- i \frac{\pi}{4}J_x^2\right) &= e^{-i\frac{j\pi}{4}} \prod_{\substack{a_1 < a_2, \\ a_3 < a_4}}^{2j} \left[\mathds{I}^{\otimes 2j} \cos\left(\frac{\pi}{8}\right) - i \sigma_z^{(a_1)} \sigma_z^{(a_2)} \sin\left(\frac{\pi}{8}\right)\right] \left[\mathds{I}^{\otimes 2j} \cos\left(\frac{\pi}{8}\right) - i \sigma_x^{(a_3)} \sigma_x^{(a_4)} \sin\left(\frac{\pi}{8}\right)\right].
\end{align}
The Pauli matrices have eigenvalues $\pm 1$. Therefore, the eigenvalues of the above tensor-product operator are $\lbrace e^{i\pi/4}, e^{-i\pi/4}\rbrace$ each with multiplicity $2^{2j-1}$. Using Caley-Hamilton theorem, we can calculate the sixth power of the above operator as follows:
\begin{align}
    {\left[\exp\left(- i \frac{\pi}{4}J_z^2\right) \exp\left(- i \frac{\pi}{4}J_x^2\right)\right]}^6 = \frac{1}{2} e^{-i\frac{3\pi}{4}2j} \left[\mathds{I}^{\otimes 2j} + {(-1)}^j {\sigma_y}^{\otimes 2j} + i {\sigma_z}^{\otimes 2j} + i {\sigma_x}^{\otimes 2j} \right].
\end{align}
Since the tensor product of Pauli matrices satisfy $\left[\sigma_a^{\otimes 2j}, \sigma_b^{\otimes 2j}\right] = 0$ for even-$2j$ and $a,b \in \lbrace x,y,z \rbrace$, the operator $\exp\left(- i \frac{k'}{2j}J_x^2\right)$ commutes with the above operator. Therefore, we get
\begin{align}
    \mathcal{U}^{12} = \; \frac{1}{2} e^{-i\frac{3\pi}{4}2j} \left[\mathds{I}^{\otimes 2j} + {(-1)}^j {\sigma_y}^{\otimes 2j} + i {\sigma_z}^{\otimes 2j} + i {\sigma_x}^{\otimes 2j} \right] \cdot {\left(\gamma^{\otimes 2j}\right)}^{12}.
\end{align}
Interestingly, the Floquet operator $\mathcal{U}^{12}$ is independent of $k_\theta$. This can also be observed in the corresponding Husimi distribution (see \cref{suppfig:husimi_even_jpiby4} and \cref{suppfig:husimi_even_qkt_jpiby4}). Again, squaring the above operator, we get
\begin{align}
    \mathcal{U}^{24} = \; - {\sigma_y}^{\otimes 2j} \implies \mathcal{U}^{48} = \mathds{I}^{\otimes 2j}.
\end{align}
This result is used in the main text.

\end{document}